\documentclass{aa}

\usepackage{graphicx}
\usepackage{natbib}

\newcommand{\casq}{\mbox{$c_{\mathrm{A}}^2$}}
\newcommand{\cssq}{\mbox{$c_{\mathrm{s}}^2$}}
\newcommand{\etaa}{\mbox{$\eta_{\mathrm{A}}$}}
\newcommand{\cv}{\mbox{$\overline{c}_\mathrm{A\!}$}}

\begin{document}

\title{Rayleigh-Taylor instability in partially ionized compressible plasmas; one
fluid approach}

\author
{A. J. D\'\i az \inst{1,2}, E. Khomenko \inst{1,2} \& M. Collados \inst{1,2} }
\institute{Instituto de Astrof\'{\i}sica de Canarias, 38205,
C/ V\'{\i}a L{\'a}ctea, s/n, La Laguna, Tenerife, Spain \and
Departamento de Astrof\'{\i}sica, Universidad de
La Laguna, 38205, La Laguna, Tenerife, Spain
\\ \email{tdiaz@iac.es; khomenko@iac.es; mcv@iac.es}}

\date{Received / Accepted  }

\authorrunning{D\'\i  az, Khomenko and Collados}
\titlerunning{RTI in partially ionized plasmas}

\abstract{}
{We study the modification of the classical criterion for the linear onset
and growth rate of the Rayleigh-Taylor instability (RTI) in a partially ionized
(PI) plasma in the one-fluid description, considering a generalized induction
equation. }
{The governing linear equations and appropriate boundary conditions,
including gravitational terms, are derived and applied to the case of the RTI
in a single interface between two partially ionized plasmas. The boundary
conditions lead to an equation for the frequencies in which some of them have
positive complex parts, marking the appearance of the RTI. We study the
ambipolar term alone first, extending the result to the full induction
equation later.}
{The configuration is always unstable because of the presence of a neutral
species. In the classical stability regime the growth rate is small, since
the collisions prevent the neutral fluid to fully develop the RTI. For
parameters in the classical instability regime the growth rate is lowered,
but for the considered theoretical values of the collision frequencies and
diffusion coefficients for solar prominences the differences with the
compressible MHD case are small.}
{PI modifies some aspects of the linear RTI instability, since it takes into
account that neutrals do not feel the stabilizing effect of the magnetic
field. For the set of parameters representative for solar prominences, our
model gives the resulting timescale comparable with observed lifetimes of RTI
plumes.}

\maketitle

\section{Introduction}

One of the well-known fluid instabilities that has been applied widely in many
different astrophysical contexts is the Rayleigh-Taylor instability (RTI for
short), which appears when a lighter fluid supports or accelerates a heavier
one. We can cite as examples the RTI in planetary nebulas \citep{babbz04},
supernova explosions \citep{fam91}, acretion disks \citep{wn83}, relativistic
jets \citep{mm13}, the evolution of the inner layers of red giants
\citep{edl06}, formation of hydrogen clouds in the local bubble \citep{bfe00} or
the solar atmospheric flux tubes \citep{p79} or prominences \citep{imsy05}, for
citing some topics among the vast literature regarding this instability in
astrophysical plasmas.

The starting point of these studies is the classical hydrodynamic instability,
with the addition of a magnetic field. To study the Rayleigh-Taylor instability
in Magnetohydrodynamics (MHD), it is better to use the divergence-free velocity
condition (which is the type of movement more likely to produce instabilities,
since the energy is not wasted in compressing the plasma). Hence, the dispersion
relation for the two fluids with densities $\rho_1$ and $\rho_2$ laying one
below the other with a magnetic field parallel to the contact surface is
\citep{chandrasekhar, Priest82},
\begin{equation}
\omega^2=  - g k \frac{\rho_2-\rho_1}{\rho_1+\rho_2} + \frac{2 B_0^2 k_x^2}
{\mu (\rho_1+\rho_2)},
\label{RTI}
\end{equation}

\noindent
where $B_0$ is the magnetic field strength and  $k$ is the modulus of the
wavenumber, $k_x$ is the component of the wavenumber along the magnetic field
direction and $k_y$ the component across the field, $g$ is the acceleration of
gravity, and $\mu$ is the magnetic permittivity. The hydromagnetic case is
recovered if no magnetic field is considered, which means that the configuration
is unstable if $\rho_2 > \rho_1$. The magnetic field stabilizes perturbations
down to a critical value of $k_x$ along its direction,
but the field cannot stabilize perturbations
across the field ($k_x=0$), no matter how strong it might be.

Other effects, such as compressibility, viscosity, tension forces, relativistic
corrections, magnetic fields perpendicular to the surface or other types of
stratification and forces, may introduce corrections to the classical stability
criterion and linear growth rate of the RTI. The leading one is normally the
compressibility, which must be taken into account in most of the applications.
There have been many studies of the effect of compressibility in the magnetic
RTI \citep[see for example][and references
therein]{v61,shivamoggi82,bb83,rtb04,livescu04,shivamoggi08,lb08}. These
studies show that the compressibility has mainly a stabilizing influence by
lowering the linear growth rates,
although the
stability threshold that appears in Eq.~\ref{RTI} is not modified. On the other
hand, even considering only this extra effect complicates greatly the solution,
so simple relations such as Eq.~\ref{RTI} are no longer obtained.

Prominences are a very likely candidate to display the RTI in the solar
atmosphere, since they are composed of cool and dense plasma surrounded by much
lighter coronal plasma \citep[see the reviews][]{lhvkpgsk10,mkbsa10}. The
magnetic field plays a key role in the structure and dynamics of prominences,
and hence the RTI must be studied in the context of the MHD theory. Using
computational techniques allow us to solve directly the MHD partial differential
equations and study the non-linear regime of the RTI.  These numerical studies
of RTI \citep[see for example][]{jns95,ahl07,sg07}
agree with the results of the linear theory and show some new interesting
features, such as the enhancement of the growth of bubbles and fingers in the
non-linear regime across the field, since it prevents secondary
Kelvin-Helmholtz instabilities and mixing between the fluids. This non-linear
phase agrees qualitatively with the observations of turbulent plumes and
bubbles in prominences
\citep{imsy05,bsstto08,hsfslka08,rbftt10,bshsttlo10,bthblsstt11} and has even
been used to infer plasma properties from observational features \citep{hht12}. The
presence of the RTI has also been confirmed numerically in more general
prominence models with 3D geometry \citep{imsy06,hisb11,hbis12,hisb12}.


Prominences have also another feature that can be relevant for the RTI because
of their physical properties, namely partial ionization. Since prominences are
relatively cool and dense objects, their plasma is expected to be partially
ionized \citep{pv02,gka07,lhvkpgsk10,zkr11}. The ionization degree of the
prominence plasma has not been directly measured with accuracy, but with the
typical physical properties of the plasma the plasma is neither in a completely
ionized state or is a neutral gas. This partially ionized prominence plasma can
no longer be described with 1-fluid ideal MHD. \citet{cv89} studied the magnetic
RTI in a two fluid model without considering the full dynamics of the neutral
fluid (which was only subject to collisions) and concluded that the stability
threshold is not changed, but the growth rate is lowered. A two-fluid
description with a neutral and an ion-electron fluids coupled only by
ion-neutral collision (without electron collisions and diffusive terms in the
induction equation) was considered in \citet{sdgb12} for the Kelvin-Helmholtz
instability  due to shear flow at an interface between two partially ionized
plasmas and in \citet{dsb12} for the RTI in a similar setup. The conclusion in
\citet{dsb12} was that the instability threshold of the RTI is modified, so the
configuration with heavier fluid on the top of lighter fluid is always unstable,
but that the linear growth rate is substantially reduced depending on the
parameters. This approach has also been used by \citet{syk13} in the context of
the RTI in the borders of the local bubble, where partial ionization also plays
a relevant role.

Our main objective in this paper is to study the stability threshold and the
linear growth rate of the RTI in a plasma described in a single-fluid model,
taking into account the partial ionization effects in the form of a generalized
induction equation and considering all the types of collisions and forces
present and the diffusive terms in the induction equation.
We develop a rather general formulation, with an aim of its global
applicability different astrophysical situations. After this general
formulation has been set, we particularize to the case of solar prominence
and consider the importance of different terms in the induction equations
(many of them have been neglected in the previous works) on the development
of the RTI in this environment.
We compare our results with those from the simple two-fluid approach in
\citet{dsb12} and try to understand the linear phase of the RTI in a
partially ionized plasma to use it as a comparison with more complicated
computational models with geometrically complex geometry and the non-linear
stages.


\section{Multifluid equations for partially ionized plasma}

\subsection{Fluid equations for each species}

The general transport equations for a multi-component plasma can be derived from
the Boltzmann kinetic equation, taking into acount general properties of he
collisions terms  \citep{Braginskii, Bittencourt, Balescu}. The most common form
of the MHD theory can only be applied to totally ionized plasma, where the
different species are completely coupled dinamically and thermally by
collisions, so partial ionization cannot be described. The first extension from
MHD in the presence of neutrals for strong collisional coupling is to consider
only the modifications due to collisions in the generalized Ohm's law and energy
transport \citep{Braginskii,karh04,fobk07}, assuming a strong thermal coupling
and neglecting the transport coefficients. Another way of including partial
ionization effects is to use a multi-fluid treatment, in which ions and
electrons are considered together as an ion-electron fluid (due to their strong
electromagnetic coupling), and neutrals are
considered separately with as many neutral species as one wishes to include
\citep[see, e.g.,][]{zkr11,zkr11b,sag12}. The two-fluid approach (a hydrogen
plasma only considered) was used in \citet{dsb12} for studing the RTI, with the
additional neglection of other diffusive terms in the induction equation.
Here we take the single fluid approach by combining the equations for each
species in a single fluid equation, and obtaining a generalized form of the MHD
equations valid for PI plasmas.

We proceed with the derivation of the generalized one fluid equations starting from
the macroscopic equations for each species. In the following expressions the
subscripts i, n and e stand for ions, neutrals and electrons, respectively. The
momentum equation for each species is

\vspace{-1mm}
\noindent
\begin{eqnarray}
\rho_e \left( \frac{\partial {\bf v}_e}{\partial t} + {\bf v}_e \cdot \nabla
	{\bf v}_e \right) &=& -\nabla p_{e} - e n_ e \left( {\bf E} +
	{\bf v}_e \times {\bf B} \right) \nonumber \\
	&+& \rho_e {\bf g} + {\bf R}_e, \nonumber \\
\rho_i \left( \frac{\partial {\bf v}_i}{\partial t} + {\bf v}_i \cdot \nabla
	{\bf v}_i \right) &=& -\nabla p_{i} + e n_ i \left( {\bf E} +
	{\bf v}_i \times {\bf B} \right) \nonumber \\
	&+& \rho_i {\bf g} + {\bf R}_i, \nonumber \\
\rho_n \left( \frac{\partial {\bf v}_n}{\partial t} + {\bf v}_n \cdot \nabla
	{\bf v}_n \right) &=& -\nabla p_n + \rho_n {\bf g} + {\bf R}_n,
\label{3fluid_eqs}
\end{eqnarray}
where ${\bf v}_i$, ${\bf v}_e$ and ${\bf v}_n$ are the velocity of the ion,
electron and neutral fluid, respectively, $p_{i}$, $p_{e}$ and $p_n$ are the
pressure of the ion, electron and the neutral fluid, respectively, $\rho_i$,
$\rho_e$ and $\rho_n$ are the ion, electron and neutron densities, respectively,
and $R_i$, $R_e$ and $R_n$ are the momentum transfer terms due to collisions for
ions, electrons and neutrals, respectively. We also define the total density
$\rho=\rho_n+\rho_i$, the neutral fraction  $\xi_n= \rho_n/\rho$ and the ion
fraction $\xi_i= \rho_i/\rho$, with $\xi_n+\xi_i=1$. Hence, the parameter
$\xi_n$ indicates the ionization degree, from $\xi_n=0$ for a fully ionized
plasma to $\xi_n=1$ for a neutral gas. We have  assumed the non diagonal terms
of the pressure tensors to be negligible and the diagonal terms to be equal, so
the pressure tensor becomes isotropic and can be represented by the scalar
pressure. The elastic collision term of each species is approximated as
\citep{Braginskii,Bittencourt}
\begin{equation}
{\bf R}_{\alpha} =
-\rho_{\alpha}\sum_{\beta}\nu_{\alpha\beta}({\bf v}_{\alpha} -
{\bf v}_{\beta})
\end{equation}

\noindent
where $\nu_{\alpha\beta}$ is the collisional frequency of species $\alpha$ with
particles of species $\beta$.
So, for our three species they become:

\vspace{-1mm}
\noindent
\begin{eqnarray}
{\bf R}_e &=& -\rho_e(\nu_{ei}({\bf v}_e - {\bf v}_i) +
	\nu_{en}({\bf v}_e - {\bf v}_n)) \nonumber \\
	&=& \rho_e \frac{1}{n_i e}{\bf J}(\nu_{ei}+\nu_{en}) -
	\rho_e \nu_{en} {\bf w} \nonumber \\
{\bf R}_i &=& -\rho_i(\nu_{ie}({\bf v}_i - {\bf v}_e)  +
	\nu_{in}({\bf v}_i - {\bf v}_n)) \nonumber \\
	&=& - \rho_i \frac{1}{n_i e} \nu_{ie} {\bf J} -
	\rho_i \nu_{in} {\bf w} \nonumber \\
{\bf R}_n &=& -\rho_n(\nu_{ni}({\bf v}_n - {\bf v}_i)  +
	\nu_{ne}({\bf v}_n - {\bf v}_e)) \nonumber \\
	&=& - \rho_n \frac{1}{n_i e} \nu_{ne} {\bf J} -
	\rho_n (\nu_{ni}+\nu_{ne}) {\bf w},
\end{eqnarray}

\noindent
with ${\bf w}={\bf v}_i - {\bf v}_n$ the diffusion velocity of ions with respect to
neutrals and ${\bf J}=e n_e ({\bf v}_i - {\bf v}_e)$ the total current density.  This
form of the friction momentum transfer does not alter the one-fluid equations of mass
conservation and momentum conservation from their MHD counterparts when written in
terms of the overall velocity of the plasma, ${\bf v}=\sum_{\alpha=i,e,i} (\rho_\alpha
{\bf v}_\alpha)/\rho \approx  \xi_i {\bf v}_i + \xi_n {\bf v}_n$. However, the energy
equation has to take into account the currents arising from these non-MHD terms, and
an additional equation for the evolution of the magnetic field is also required to
close the system, namely a generalized induction equation.

\subsection{Equation for the diffusion velocity}

To obtain the momentum equation for the diffusion velocity ${\bf w}$ we proceed
as follows. The momentum equation for electrons and ions are added up,
neglecting the electron inertial terms because of the small electron mass
compared to the other species. We are not neglecting the electron gravity
compared with the ion gravity yet ($\rho_e{\bf g}$ compared to $\rho_i{\bf g}$),

\vspace{-1mm}
\noindent
\begin{eqnarray}
\rho_i \left( \frac{\partial {\bf v}_i}{\partial t} \right. \!\!&+&\!\! \left.
	\rule{0mm}{5.2mm}
	{\bf v}_i \cdot \nabla {\bf v}_i \right) = {\bf J} \times {\bf B}  +
	(\rho_i+\rho_e) {\bf g} - \nabla (p_i + p_e)  \nonumber \\
	&-& \!\! \alpha_n {\bf w} + \rho_n \nu_{ne} \frac{1}{n_i e}
	{\bf J} \nonumber \\
\rho_n \left( \frac{\partial {\bf v}_n}{\partial t} \right. \!\!&+&\!\! \left.
	\rule{0mm}{5.2mm} {\bf v}_n \cdot \nabla {\bf v}_n \right) =  \rho_n
	{\bf g} - \nabla p_n + \alpha_n {\bf w} \nonumber \\
	&-& \rho_n \nu_{ne} \frac{1}{n_i e} {\bf J},
\end{eqnarray}

\noindent
with the definition for the coefficient of friction between the plasma and the
neutral gas
\begin{equation}  \label{alphan}
\alpha_n=\rho_n (\nu_{ni}+\nu_{ne})= \rho_i \nu_{in} + \rho_e \nu_{en}.
\end{equation}

\noindent
Now we add the upper equation multiplied by $\xi_n$ and lower
equation, multiplied by $-\xi_i$. The result is:

\vspace{-1mm}
\noindent
\begin{eqnarray}
\xi_i\xi_n\rho \left( \frac{D_i {\bf v}_i}{Dt} -\frac{D_n {\bf v}_n}{Dt}
\right ) = \xi_n \left[{\bf J} \times {\bf B} \right] - {\bf G} -
\alpha_n {\bf w} \nonumber \\
+ \rho_n \nu_{ne} \frac{1}{n_i e} {\bf J} + \xi_n \rho_e {\bf g},
\label{dif_eq1}
\end{eqnarray}

\noindent
where
\begin{equation}
\frac{D_\alpha}{D t}=\frac{\partial}{\partial_t} + v_\alpha \cdot \nabla.
\end{equation}

\noindent
We have taken into account that $\xi_i+\xi_n=1$, thus the gravity terms for ions
and neutrals cancel out and the electron term might become relevant. This
gravitional term is in fact similar to those of the electron inertia that have
been already neglected, but we keep it here to check its magnitude, specially
since the RTI is driven by gravity so it is important to assure its effect as
much as possible. Regarding the pressure gradients, we have introduced the new
PI pressure terms following \citet{Braginskii} as
\begin{equation} \label{G_brag}
{\bf G}= \xi_n \nabla (p_i + p_e) - \xi_i \nabla p_n.
\end{equation}

\noindent
We still have the ion and neutral inertia terms in Eq.~(\ref{dif_eq1}). These
are neglected on the basis of the following argument. We can express the total
derivative in terms of the diffusion velocity

\noindent
\begin{eqnarray}
\left(\frac{D_i {\bf v}_i}{Dt} - \frac{D_n {\bf v}_n}{Dt} \right) =
 \frac{\partial {\bf w}}{\partial t} +  (\xi_n-\xi_i)  {\bf w} \cdot \nabla
 {\bf w} +  \nonumber \\
 {\bf w} \cdot \nabla  {\bf v} + {\bf v} \cdot \nabla  {\bf w}.
	\approx \frac{\partial {\bf w}}{\partial t}.
\end{eqnarray}

\noindent
In a linear regime all the advection terms in this equation are second order
effects, and the remaining time derivative of ${\bf w}$
can be neglected when compared with the friction terms,
which are of the order of ${\bf w}/\tau_\mathrm{col}$, with
$\tau_\mathrm{col} \sim 1/ \nu_{\alpha \beta}$ being the characteristic
timescale related to collisions.

Taking all the aforementioned simplifications into
account, we obtain an expression for the diffusion velocity between ions and
neutrals,
\begin{equation}
{\bf w} = \frac{\xi_n}{\alpha_n} {\bf J}
\times {\bf B} - \frac{ {\bf G} }{\alpha_n} +
\rho_n \nu_{ne} \frac{1}{n_ie\alpha_n} {\bf J}
+ \frac{\xi_n \rho_e}{\alpha_n} {\bf g}.
\label{eq_w}
\end{equation}

\noindent
From this expression we can see that the ion and neutral fluids do not follow each
other exactly, which raises additional dissipative effects.
On the other hand, by
neglecting the inertial terms we have obtained an explicit expression for the
diffision velocity in terms of other variables, assuming that collissions lead to
the terminal values of ${\bf w}$ given by Eq.~(\ref{eq_w}) much faster than the
evolution of the remaining variables. Thus, the velocities of the ion and neutral
species do not longer appear in the equations and the diffusion velocity
${\bf w}$ can be computed from the single-fluid variables.

\subsection{Induction Equation}

To proceed further we need a generalized Ohm's law and an induction equation to
obtain an equation for the magnetic field evolution. These are obtained from the
momentum equation for electrons in Eq.~\ref{3fluid_eqs}, neglecting again their
inertial terms. We obtain after some algebra

\vspace{-1mm}
\noindent
\begin{eqnarray}
{\bf E} + {\bf v}\times{\bf B} &=&
-\xi_n \, {\bf w}\times{\bf B} + \frac{1}{n_i e} {\bf J}\times{\bf B} \nonumber \\
&-& \frac{\nabla p_e}{e n_e} + \alpha_e \frac{1}{n_i^2e^2} {\bf J} -
\frac{\rho_e\nu_{en}{\bf w}}{en_e} + \frac{\rho_e}{n_e e} {\bf g},
\label{ohmlaw}
\end{eqnarray}

\noindent
with the definition $\alpha_e=\rho_e (\nu_{ei}+\nu_{en})$.
We then substitute the expression for the diffusion velocity in Eq.~(\ref{eq_w})
and insert the result in Faraday's law, obtaining

\vspace{-1mm}
\noindent
\begin{eqnarray}
\frac{\partial{\bf B}}{\partial t}  &=&  {\bf \nabla}\times \left[
({\bf v}\times{\bf B}) -  \frac{{\bf J}}{\sigma} -\left(
\frac{1-2\varepsilon\xi_n}{en_e}{\bf J}\times{\bf B}\right) \right. \nonumber \\
&+& \left.  \left( \frac{\xi_n^2}{\alpha_n}({\bf J}\times{\bf B})
\times{\bf B}\right)  - \left( \frac{\varepsilon{\bf G} -
\nabla p_e}{en_e} \right) \right. \nonumber \\
&-& \left. \left(\frac{\xi_n}{\alpha_n}{\bf G}\times{\bf B} \right)
-\left( \frac{\rho_e}{e n_e} (1+\xi_n \varepsilon)
{\bf g} \right. \right. \nonumber \\
&+& \left. \left. \frac{\xi_n^2 \rho_e}{\alpha_n} {\bf g} \times {\bf B}  \right)
\right]
\label{eq_induction}
\end{eqnarray}

\noindent
with the definitions of $\varepsilon=\rho_e\nu_{en}/\alpha_n$ (which is a small
parameter) and the Ohmic conductivity $\sigma = (en_e)^2 /( \alpha_e
-\varepsilon^2 \alpha_n)$. In this equation, the terms on the right hand side
are: ideal MHD induction term, Ohmic term, Hall term, ambipolar term,
generalized battery term (which includes a part already present in 
plasmas and a part depending on partial ionization by means of ${\bf G}$), a
${\bf G} \times {\bf B}$ term of perpendicular currents caused by these pressure
gradients (similar to the Hall term with currents) and gravity terms (again with
a similar ${\bf g} \times {\bf B}$ part also included). We can define  the
coefficients in the different terms as

\vspace{-1mm}
\noindent
\begin{eqnarray}
\eta & = & \frac{1}{\sigma\mu} = \frac{\alpha_e -
(\rho_e\nu_{en})^2/\alpha_n}{(en_e)^2\mu}, \nonumber \\
\eta_\mathrm{H} & = & \frac{1-2\varepsilon\xi_n}{en_e\mu} B_0, \nonumber \\
\eta_\mathrm{A} & = & \frac{\xi_n^2}{\alpha_n\mu} B_0^2,  \nonumber \\
\chi_\mathrm{p} & = & \frac{\xi_n}{\alpha_n}, \nonumber  \\
\chi_\mathrm{g} & = & \frac{\xi_n^2 \rho_e}{\alpha_n} ,
\label{eta_defs}
\end{eqnarray}

\noindent
with $\eta$, $\eta_\mathrm{H}$, $\eta_\mathrm{A}$, $\chi_\mathrm{p}$ and
$\chi_\mathrm{g}$ being the ohmic diffusivity, Hall diffusivity, ambipolar
diffusivity and coefficients related to the battery and gravity
terms, respectively. Assuming that the acceleration of gravity is uniform, the
curl of the pre-last term in  Eq.~\ref{eq_induction} vanishes, and the induction
equation is finally written as

\vspace{-1mm}
\noindent
\begin{eqnarray}
\frac{\partial{\bf B}}{\partial t}  &=&  {\bf \nabla}\times \left[ \rule{0mm}{3.5mm}
({\bf v}\times{\bf B}) -  \eta \nabla \times {\bf B} -
\eta_\mathrm{H} \left(\nabla \times {\bf B} \right) \times{\bf B}/B_0
\right. \nonumber \\
&+& \left. \rule{0mm}{3.5mm} \eta_\mathrm{A} \left\{ \left( \nabla \times {\bf
B} \right) \times{\bf B} \right\} \times{\bf B} /B_0^2  - \left(\varepsilon{\bf
G} -
\nabla p_e \right) /(en_e) \right. \nonumber \\
&-&  \left. \rule{0mm}{3.5mm}\chi_\mathrm{p} {\bf G}\times{\bf B}
- \chi_\mathrm{g} {\bf g} \times {\bf B} \right],
\label{ind_etas}
\end{eqnarray}

\noindent
in which we used Ampere's law (neglecting Maxwell's displacement current)
to eliminate the current density in terms of the magnetic field,
${\bf J} = \nabla \times {\bf B}/\mu$. Eq.~\ref{ind_etas} is a very general form of
the induction equation in the one fluid description of partial ionized plasmas, and
it is in fact a generalization of the well-known generalized induction equation in
classical textbooks \citep[see for example][]{Braginskii} with all the pressure
gradient and gravity terms included and the expressions of the diffusion
coefficients.

The formulation of the induction equation in Eq.~(\ref{ind_etas}) allows
for the very general analysis, that can be useful in a broad context of
astrophysical plasmas. Some of the terms are a priory expected to be smaller
than others (as those related to the electron mass), but others can not be
ruled out just from general considerations. Below we will discuss their
importance for the case of the parameters appropriate for solar prominences.
We next describe the plasma and magnetic field configuration used to study
the RTI in this environment in Sect.~\ref{sect_refconf}  and then explore in
Sect.~\ref{sect_ambip} the effect of the leading term under these conditions,
namely the ambipolar diffusion $\eta_\mathrm{A} \left\{ \left( \nabla \times
{\bf B} \right) \times{\bf B} \right\} \times{\bf B} /B_0^2 $. Then, we
consider the full induction equation in Sect.~\ref{sect_fullind} to test the
magnitude of the remaining terms and finish discussing the results and
drawing our conclusions.

\section{Reference configuration} \label{sect_refconf}

Since we are aiming to obtain some extensions to the well known formula in
Equation~(\ref{RTI}) we restrict the analysis to the simple configuration of a
contact surface following the classical analysis in
\citep{chandrasekhar,DrazinReid,Priest82}, amenable to analytical solutions. We
use this configuration to study the RTI in prominence threads, specially for
choosing the values of the equilibrium and perturbation parameters,
but the method developed in this paper is general and can be applied to other
astrophysical situations which involve the RTI in PI plasmas.

The reference configuration consists of two regions filled with uniform
plasmas composed of ions, electrons and neutrals separated by a contact surface
at $z=0$. We use Cartesian coordinates and denote the quantities in the plasma
below the discontinuity ($z<0$) with a subscript 1 and those in the plasma above
the discontinuity ($z>0$) with a subscript 2. The magnetic field permeating the
plasma is uniform and tangent to the discontinuity, so ${\bf B}=B_0 \hat{x}$,
while gravity is perpendicular to it, so ${\bf g}=-g \hat{z}$. The whole
configuration is invariant in the $x$ and $y$-directions.

\begin{figure}[h]
  \center{\resizebox{\hsize}{!}{\includegraphics{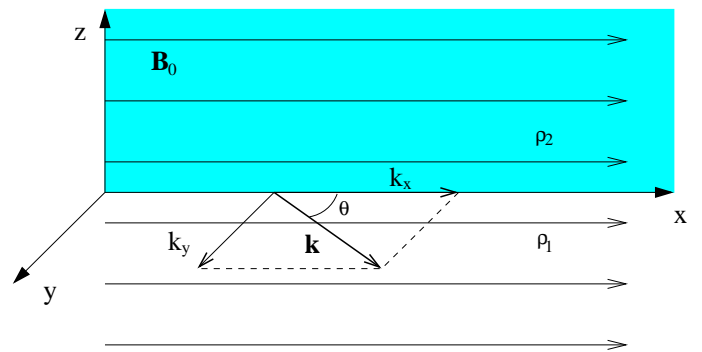} }}
\vspace{-2mm}
\caption{Sketch of the equilibrium configuration used in the analysis of this work.
The equilibrium state is a contact surface between two regions filled uniformly
with plasma  having different properties, with the lower quantities labelled as
``1'' and the upper ones as ``2''. The magnetic field is uniform and directed along
the $x$-axis, while the whole configuration is invariant in the $x$ and
$y$-directions.
}
\label{sketch}
\end{figure}

In absence of hydrostatic pressure gradients or flows, the plasma described in
the previous paragraph is not in
equilibrium, since nothing counteracts the gravity force. We are not interested in
the overall equilibrium, only in the local region where the instability is
triggered. More precisely, the equilibrium pressure gradient $\nabla p_0$ is
related to the gravitational scale height, while the perturbed quantities vary on
a much shorter spatial scale.
Hence, we assume that all the plasma magnitudes (namely the density, the
pressure and the ionization degree) are constant in each zone. Pressure balance
along the discontinuity demands that the total pressure for each species must be
equal in each side, and since the magnetic field is assumed to be  uniform this
means the gas pressures are equal in each side ($p_1=p_2$).
On the other hand, the temperature, density and ionization degree on each region
are parameters of our model. Since we are interested in studing the RTI we
assume that $\rho_2>\rho_1$. No ionization-recombination processes are included,
so the ionization degree in each region remains constant. Note also that the
plasma beta $\beta=c_\mathrm{s2}^2/c_\mathrm{A2}^2=
c_\mathrm{s1}^2/c_\mathrm{A1}^2 =\gamma \mu p_2 / B_0^2$ is then constant all
over the domain.

Another important simplification in this particular configuration is that
we can neglect the variations of the coefficients in Eq.~(\ref{eta_defs}) during
the evolution of the instability.
The equilibrium field satisfies $\nabla \times {\bf
B}_0 =0$, so there are no currents in the reference state and the only
contribution of the diffusive terms to the first-order induction equation are
those with the coefficients calculated in the reference configuration.

Finally, using this reference configuration the classical instability criterion
(Eq.~\ref{RTI}) can be written in the following form
\begin{equation}
\label{stab_crit}
\omega^2=2\left(\cv^2 \mathrm{cos}^2 \theta - c_\mathrm{crit}^2
\right) \, k^2,
\end{equation}

\noindent
with $\theta$ being the angle between the equilibrium magnetic field and the
wavevector ${\bf k}$ and $k$ the wavevector modulus (wavenumber). We define the
critical speed as
\begin{equation}
c_\mathrm{crit}=\left( \frac{g}{2 k} \,\, \frac{\rho_2-\rho_1}{\rho_2+\rho_1}
\right)^{1/2},
\end{equation}

\noindent
and the reduced square Alfv\'en speed as
\begin{equation}
\cv=\left( \frac{B_0}{\mu_0 (\rho_1+\rho_2)} \right)^{1/2},
\end{equation}

\noindent
with usual definition for the squared Alfv\'en speed $\casq=B_0^2/(\mu
\rho_{0})$, so $\cv^{-2}=(c_\mathrm{A1})^{-2}+(c_\mathrm{A2})^{-2}$
It is convenient then to use $\cv$ as a
parameter, and notice that in the case of a prominence with $\rho_2 \gg \rho_1$
we have $\cv \approx c_\mathrm{A2}$, so this averaged Alfv\'en speed is
approximately the Alfv\'en speed in the prominence.

\section{MHD plus ambipolar diffusion} \label{sect_ambip}

It is clear that dealing with the all the terms in Eq.~\ref{ind_etas} is very
difficult. Hence, we concentrate first in the modifications introduced in the ideal
MHD theory by the ambipolar term, which has proved to be relevant in solar
atmospheric situations \citep[see for example][and references
therein]{karh04,ahl07,kc12}. We neglect in this section all the magnetic diffusion
terms except the ambipolar one, obtaining a simple form of the induction equation,
\begin{equation}
\frac{\partial {\bf B}}{\partial t}=\nabla \times \left( \rule{0mm}{3.5mm}
{\bf v} \times {\bf B} + \eta_\mathrm{A} \left\{ \left( \nabla \times {\bf B}
\right) \times{\bf B} \right\} \times{\bf B} /B_0^2
\right).
\end{equation}

The mass and momentum conservation equations are not modified by the presence of
ambipolar diffusion. In addition, we assume an adiabatic energy equation plus
the contribution from the ambipolar diffusion term (neglecting transport terms
such as conduction, radiation of other heating sources). Hence, after deriving
the energy term corresponding to the ambipolar diffusion, our system of basic
equations is

\vspace{-1mm}
\noindent
\begin{eqnarray}
\frac{\partial \rho}{\partial t} &=& - \nabla \cdot \left( \rho {\bf v}
	\right), \nonumber \\
\rho \frac{\partial {\bf v}}{\partial t} &=& - \rho {\bf v} \cdot \nabla
	{\bf v} -\nabla p +  \frac{1}{\mu} \left( \nabla \times
	{\bf B}\right) \times {\bf B} + \rho {\bf g}, \nonumber \\
\frac{\partial {\bf B}}{\partial t} &=& \nabla \times \left( \rule{0mm}{3.5mm}
	{\bf v} \times {\bf B} + \eta_\mathrm{A}
	\left\{ \left( \nabla \times {\bf B} \right) \times{\bf B} \right\}
	\times{\bf B} /B_0^2 \right),\nonumber \\
\frac{\partial p}{\partial t} &=& - {\bf v} \cdot \nabla p
	- \gamma p \nabla \cdot {\bf v} \nonumber \\
	&\!& \!\!\!\!\!\!\!\!\! + (\gamma-1) \frac{\eta_\mathrm{A}}{\mu B_0^2}
	\left(\nabla \times {\bf B} \right) \times
	\left[ \left\{ \left( \nabla \times {\bf B} \right) \times{\bf B}
	\right\} \times{\bf B} \right],
\end{eqnarray}

\noindent
where $\gamma$ is the adiabatic index and $p=p_i+p_e+p_n$ the total scalar
pressure of the fluid.

\subsection{Linearized equations}

Next, we study linear perturbations from the uniform state. To obtain a general
formulation, we label the reference
quantities with the subscript 0 and the linear perturbations without subscript
(${\bf B}=B_0 {\bf x}+{\bf b}$). The subscript 0 can be replaced with 1 of 2
when one of the regions in which the physical domain is considered, but
otherwise, the deduction is valid for any uniform configuration.

Since no equilibrium flow is present (${\bf v}_0=0$), all the advection terms are
second order effects. No currents are present in the reference state, so the term
in the energy equation coming from the ambipolar diffusion is also a second order
effect and can be neglected in the linealized problem. Hence, we are left with a
considerably simpler system of  differential equations, namely

\vspace{-1mm}
\noindent
\begin{eqnarray}
\frac{\partial \rho}{\partial t} &=& - \rho_{0} \nabla \cdot {\bf v}, \nonumber \\
\rho_{0} \frac{\partial {\bf v}}{\partial t}  &=& -\nabla p +
	\frac{1}{\mu} \left( \nabla \times {\bf b}\right)
	\times {\bf B}_0 + \rho {\bf g},  \nonumber \\
\frac{\partial {\bf b}}{\partial t} &=& \nabla \times \left( \rule{0mm}{3.5mm}
	{\bf v} \times {\bf B}_0 + \frac{\eta_\mathrm{A}}{B_0^2} \left\{ \left(
	\nabla \times
	{\bf b} \right) \times{\bf B}_0 \right\}  \times{\bf B}_0 \right),
	\nonumber \\
\frac{\partial p}{\partial t} &=&
	- \gamma p_{0} \nabla \cdot {\bf v}.
\label{linear_eqs4}
\end{eqnarray}

\noindent
This system of equations can be reduced to only two by taking the time derivative
of the momentum equation and substituting the expressions for the density and
pressure from the continuity and energy equations, respectively, obtaining a system
of two partial differential equation for the perturbations in velocity and magnetic
field, namely

\vspace{-1mm}
\noindent
\begin{eqnarray}
\frac{\partial^2 {\bf v}}{\partial t^2} &=& \casq \left(\nabla \times
   \frac{\partial {\bf b}}{\partial t}
   \right) \times {\bf B}_0 - \left(\nabla \cdot {\bf v}\right) {\bf g} +
   \cssq \nabla \left( \nabla \cdot {\bf v} \right), \nonumber \\
\frac{\partial {\bf b}}{\partial t} &=& \nabla \times \left( \rule{0mm}{3.5mm}
	{\bf v} \times {\bf B}_0 + \frac{\eta_\mathrm{A}}{B_0^2} \left\{ \left(
	\nabla \times
	{\bf b} \right) \times{\bf B}_0 \right\}  \times{\bf B}_0 \right),
\label{linear_eqs}
\end{eqnarray}

\noindent
We have defined the squared sound speed as $\cssq=\gamma p_{0}/\rho_{0}$. The
equilibrium properties of the medium relevant for the RTI are included in  the
sound and Alfv\'en speeds, together with the ambipolar diffusivity coefficient.
In our problem is not possible to derive a single equation by eliminating the
Lorentz force term in the motion equation by using the induction equation as is
routinely done in ideal MHD \citep[see for
example][]{Roberts81,Priest82,gp04book}.

We are left with only one parameter depending on the ionization fraction, the
ambipolar diffusivity $\etaa$ (Eq.~(\ref{eta_defs}), which is calculated in the
equilibrium state and depends on the ionization degree $\xi_n$ and the neutral
friction coefficient (Eq.~\ref{alphan}),
\begin{equation}
\alpha_n= \rho_0 (1-\xi_n) ( \nu_{in} + \nu_{en} m_e/m_i).
\end{equation}

\noindent
We can neglect the term with the ratio of the electron and ion mass and use the
expression for the ion-neutral collision frequency in a plasma
\citep[see, e.g., ][]{Braginskii,sob09b},
\begin{equation}
\nu_{in}=\frac{\rho_n}{m_\mathrm{n}} \sqrt{\frac{16 k_\mathrm{B} T}
{\pi  m_\mathrm{n}}} \, \sigma_\mathrm{in},
\label{nin}
\end{equation}

\noindent
where $T$ is the temperature, $m_\mathrm{n}$ the neutron mass, $k_\mathrm{B}$ is
the Boltzmann constant and $\sigma_\mathrm{in} \approx 5 \times 10^{-19}$ m$^2$
is the collisional cross section for proton-hydrogen collisions (assuming a
hydrogen plasma). Notice that
this collision frequency is a theoretical value for hard-sphere collisions
between protons and H molecules, while there are hints that actual values
can differ from this simple calculation \citep{MitchnerKruger73,vk13}. A
strong thermal coupling is assumed, so the temperature of the different
species is the same. Hence, $ k_\mathrm{B} T/m_n = p_n/\rho_n=(2-\xi_n)^{-1}
\cssq/\gamma$ and the expression for the friction coefficient is
\begin{equation}
\alpha_n=\frac{4 \sigma_\mathrm{in}}{m_n \left( \pi \gamma \right)^{1/2}} \,
	\rho_0^2 c_\mathrm{s} \frac{\xi_n (1- \xi_n)}
	{(2-\xi_n)^{1/2}},
\end{equation}

%

\noindent
Finally, we obtain from Eq.~(\ref{eta_defs}) an equation for $\etaa$, in therms of
the equilibrium parameters in each region.
\begin{equation}
\etaa=  \frac{m_n \left( \pi \gamma \right)^{1/2}}{4 \sigma_\mathrm{in}} \,
   \frac{\xi_n (2-\xi_n)^{1/2}}{1-\xi_n} \,
   \frac{\casq}{\rho_0 c_\mathrm{s}}.
\label{eta_param}
\end{equation}

\noindent
This expression depends on the medium density. We use $\rho_0=10^{10}$ kg
m$^{-3}$ for computing this coefficient in the prominence region through the
paper, a value representative of typical densities in prominences, while the
value in the corona is adjusted taking into account the prominence-corona
contrast ratio $\rho_2/\rho_1$ used in each calculation.


\subsection{Normal mode analysis}

We consider the normal mode decomposition and write the temporal dependence of the
perturbation as $e^{-i \omega t}$. We Fourier analyze in the spatial directions
where the medium is uniform and write the perturbations as $e^{i k_x x + i k_y y}$,
with $k_x$ and $k_y$ the wavenumbers in the $x$ and $y$-directions, respectively,
and ${\bf k}=k_x \hat{x} + k_y \hat{y}$ the wavenumber parallel to the surface.

Then, we combine Eqs.~(\ref{linear_eqs}) and arrive at a system of two coupled
equations for $v_z$, the $z$-component of the velocity (normal to the surface) and
$b_x$, the $x$-component of the perturbation of the magnetic field (along the
equilibrium magnetic field)

\vspace{-1mm}
\noindent
\begin{eqnarray}
i \omega \casq (\omega^2 - k_x^2 \cssq) \frac{d b_x}{dz} - \frac{i g k_y^2 \omega^2
\casq (\omega + i \etaa k_x^2) }{\omega^2+ i k_x^2 (\omega \etaa - i \casq)} \, b_x =
\nonumber \\
\omega \cssq (\omega + i k_x^2 \etaa) \frac{d^2 v_z}{d z^2} - g \omega (\omega +
i k_x^2 \etaa) \frac{d v_z}{d z} \nonumber \\
+ \left[ \omega (\omega^2 - k^2 \cssq) (\omega + i
k_x^2 \etaa) - k_x^2 \casq (\omega^2- k_x^2 \cssq) \right] \, v_z,
\label{bx_eq}
\end{eqnarray}

\vspace{-1mm}
\noindent
\begin{eqnarray}
\frac{d v_z}{dz} &=& \etaa \left\{-k_x^2 \casq(\omega^2-k_x^2 \cssq)+ \omega
(\omega^2 - k^2 \cssq)(\omega \right. \nonumber \\
&+& \left. i k_x^2 \etaa)\right\}/\left\{(\omega^2 - k_x^2 \cssq)(\omega^2 - k_x^2
\casq
\right. \nonumber \\
&+& \left. i \omega \etaa k_x^2)\right\} \frac{d^2 b_x}{d z^2} + \left\{i (\omega
+ i k_x^2 \etaa)[-k^2 \casq (\omega^2 \right. \nonumber \\
&-& \left. k_x^2 \cssq) +\omega (\omega^2 - k^2 \cssq) (\omega + k^2 \etaa) ]
 \right\}/\left\{(\omega^2 \right. \nonumber \\
&-& \left. k_x^2 \cssq) (\omega^2 - k_x^2 \casq + i \omega \etaa k_x^2)
\right\} \, b_x.
\label{vz_eq}
\end{eqnarray}

\noindent
We can further operate these equations to obtain a single differential equation,
\begin{equation}
C_4 \frac{d^4 b_x}{d z^4} + C_3 \frac{d^3 b_x}{d z^3} +
C_2 \frac{d^2 b_x}{d z^2} + C_1 \frac{d b_x}{d z} +
C_0 \, b_x=0,
\label{mdc}
\end{equation}

\noindent
with the following definitions for the coefficients,

\vspace{-1mm}
\noindent
\begin{eqnarray}
C_4 &=& \omega \cssq \etaa, \nonumber \\
C_3 &=& - g \omega \etaa, \nonumber \\
C_2 &=& i \casq (\omega^2 - k_x^2 \cssq) + \omega \left[ \omega^2
	\etaa + \cssq (i \omega - 2 k^2 \etaa) \right], \nonumber \\
C_1 &=& -i g \omega (\omega + i k^2 \etaa), \nonumber \\
C_0 &=& i \left[ -k^2 \casq (\omega^2 - k_x^2 \cssq) \right. \nonumber \\
    &\,& + \left. \omega (\omega^2 -
	k^2 \cssq) (\omega + i k^2 \etaa )\right].
\end{eqnarray}

\noindent
This ordinary differential equation is valid in each zone, with the equilibrium
quantities $\casq$, $\cssq$ and $\etaa$ with subscripts 1 or 2 when applied to
the two regions above and below the contact surface at $z=0$, respectively.

Finally, we need the boundary conditions to match the solutions at the boundary
$z=0$. In ideal MHD the continuity of the normal component of the
velocity perturbation and the continuity of the total pressure are enough, plus the
contribution of gravity from the momentum balance at the boundary. However, in this
particular problem additional constraints are necessary. We derive these conditions
by integrating Eqs.~\ref{linear_eqs4} across the
surface $z=0$ and doing the limit of infinitesimal integration volume. We
obtain only four independent jump relations, namely

\vspace{-1mm}
\noindent
\begin{eqnarray}
\left[ \rho_{0} \cssq v_z \right] =0, \nonumber \\
\left[ \rho_{0} (i \omega \casq b_x - i k_x \cssq v_x -i k_y \cssq v_y +
	\cssq v_z' - g v_z)\right] = 0, \nonumber \\
\left[ i k_x \etaa b_z -v_z +\etaa b_x' \right]=0, \nonumber \\
\left[ \etaa b_x \right]=0,
\label{bc2_vs}
\end{eqnarray}

\noindent
where the prime represents a derivative on the $z$-direction and
$[X]=X_2(0^+)-X_1(0^-)$ stands for the jump of the quantity $X$ across $z=0$.
Expressing the components of the perturbed velocity in terms of $b_x$ and $v_z$
we obtain the set of jump relations required for our system (since $v_z$
requires the integral of $b_x$). The first relation is just the typical boundary
condition for the velocity perturbation (since $\rho_0 \cssq= \gamma p_0$ is
equal in both sides due to the equilibrium pressure balance), and the second is
related to the momentum balance (the first term is related to the magnetic
pressure, the next three to the gas pressure and the last one to the gravity
force), but the remaining two conditions come from the new terms from the
induction equation, which for $\etaa=0$ do not give any additional information.
In terms of $v_x$ and $b_z$ our final set of boundary conditions is

\vspace{-1mm}
\noindent
\begin{eqnarray}
\left[ v_z \right] &=&0, \nonumber \\
\left[ i \omega \rho_0 \frac{\casq (\omega^2-k_x^2 \cssq) + \omega \cssq (\omega +
	i k^2 \etaa)}{\omega^2 - k_x^2 \cssq} \, b_x  \right. &\,& \nonumber \\
	+ \left. \rho_0 g v_z +
	\frac{\omega^2 \cssq \etaa b_x''}{\omega^2 - k_x^2 \cssq} \right] &=& 0,
	\nonumber \\
\left[ \frac{\etaa b_x' - v_z}{\omega + i k_x^2 \etaa} \right] &=& 0, \nonumber \\
\left[ \etaa b_x \right] &=& 0,
\label{bc}
\end{eqnarray}

We
use the standard definition for the linear growth rate of the instability,
Im($\omega$).
%
%
Hence, Im$(\omega) > 0$ is related to unstable modes, while
Im$(\omega) < 0$ marks a damping in the wave.
We also define the density contrast $\rho_2/\rho_1$ and
$\beta=c_\mathrm{s2}^2/c_\mathrm{A2}^2= c_\mathrm{s1}^2/c_\mathrm{A1}^2 =\gamma
\mu p_2 / B_0^2$ as a measure of the magnetic field strength compared with the
pressure terms.

\subsection{Fully ionized plasma} \label{MHD_limit_sect}

Before dealing with the full problem, we check the known limit of ideal MHD.
This is achieved by considering a fully ionized plasma and letting $\etaa \to
0$. In this case, the equations are highly simplified, and Eq.~(\ref{mdc}) just
becomes
\begin{equation}
\frac{d^2 b_x}{dz^2} - \frac{g \omega^2}{\Omega}
\frac{d b_x}{dz} + \frac{\omega^2-(k_x^2+k_y^2)\Omega}{\Omega} \, b_x=0,
\label{MHD_limit}
\end{equation}

\noindent
with $\Omega=\omega^2 (\casq+\cssq) - k_x^2 \casq \cssq$.
We also obtain the relation
\begin{equation}
\frac{d v_z}{d z}=i \omega \frac{g \omega^2 b_x - \omega \Omega b_x'}
{(\omega^2-k_x^2 \casq)(\omega^2-k_x^2\cssq)}
\end{equation}

\noindent
This differential equation describes the propagation of ideal MHD modes. For
example, inserting a solution $b_x=A e^{i k_z z}$ we
recover the well-known dispersion relation for the MHD fast and slow
modes present when gravity is not taken into account.
Moreover, by imposing $\etaa \to 0$
the boundary conditions in Eqs.~\ref{bc} reduce to the continuity of the normal
component of the velocity perturbation and the continuity of the total pressure
(plus a gravity term), with the two extra conditions either identically
vanishing or reducing to those, and thus recovering ideal MHD boundary
conditions.

We can study the linear phase regime of the compressible MHD RTI by solving
Eq.~(\ref{MHD_limit}). The solutions of its indicial (characteristic) equation
obtained after setting $b_x=A_\mathrm{ind} e^{mz}$ (with $A_\mathrm{ind}$ an
arbitrary constant) are
\begin{equation}
m_\pm=\frac{g \omega^2}{2 \Omega} \pm \left( k_x^2 + k_y^2 - \frac{\omega^4}
	{\Omega} +\frac{g^2 \omega^4}{4 \Omega^2} \right)^{1/2}.
\label{m_sc}
\end{equation}

\noindent
We need to choose the solution in each region that guarantees the perturbation
to vanish far from the discontinuity. Hence, the solution is
\begin{equation}
b_x(z)= \left\{ \begin{array}{ll}
    A_{1} \, e^{m_{1+} z}, & z < 0 , \\
    A_{2} \, e^{m_{2-} z}, & z > 0 , \end{array} \right.
\label{mhd_sol}
\end{equation}

\noindent
with $A_1$ and $A_2$ being constants. Applying the remaining boundary conditions
in Eq.~(\ref{bc}) we obtain the dispersion relation for the system

\vspace{-1mm}
\begin{eqnarray}
\rho_1 \left\{ g + \frac{\omega^2-k_x^2 c_\mathrm{A1}^2}{m_1} - \frac{g \omega^2
	(\omega^2-k_x^2  c_\mathrm{A1}^2)}{m_1 g \omega^2-m_1^2
\Omega_1}
 \right\}= \nonumber \\
\rho_2 \left\{ g + \frac{\omega^2-k_x^2 c_\mathrm{A2}^2}{m_2} - \frac{g \omega^2
	(\omega^2-k_x^2  c_\mathrm{A2}^2)}{m_2 g \omega^2-m_2^2
\Omega_2} \right\},
\label{dr_mhd}
\end{eqnarray}

There are a couple of interesting limiting cases to this expression. If we set $g=0$
we recover the solution for surface MHD waves in an interface
\citep[see, e.g.,][]{w79,Roberts81} with no instabilities, namely
\begin{equation}
\rho_1 \frac{\omega^2-k_x^2 c_\mathrm{A1}^2}{m_1}=
\rho_2 \frac{\omega^2-k_x^2 c_\mathrm{A2}^2}{m_2},
\end{equation}

\noindent
with $m_i^2=k_x^2+k_y^2+\omega^4/\Omega_i$. Another interesting limit
is the incompressible case, obtained if we set $c_\mathrm{si1} \rightarrow
\infty$ and $c_\mathrm{si2} \rightarrow \infty$. The dispersion relation is then
\begin{equation}
\rho_1 \frac{g m_1 + \omega^2-k_x^2 c_\mathrm{A1}^2}{m_1}=
\rho_2 \frac{g m_2 + \omega^2-k_x^2 c_\mathrm{A2}^2}{m_2},
\end{equation}

\noindent
with $m_1= k$ and  $m_2= -k$ from Eq.~(\ref{m_sc}) in this limit, so we can
obtain an explicit equation for the frequencies of the modes,
\begin{equation}
\omega^2= - g k \frac{\rho_{2}-\rho_{1}}{\rho_{1}+\rho_{2}} +
\frac{(\rho_{1} c_\mathrm{A1}^2 + \rho_{2} c_\mathrm{A2}^2) k_x^2}
{(\rho_{1}+\rho_{2})},
\label{clas_MRTI}
\end{equation}

\noindent
which is equivalent to the classical RTI relation in Eq.~(\ref{RTI}).

\begin{figure}[h]
  \center{\resizebox{\hsize}{!}{\includegraphics{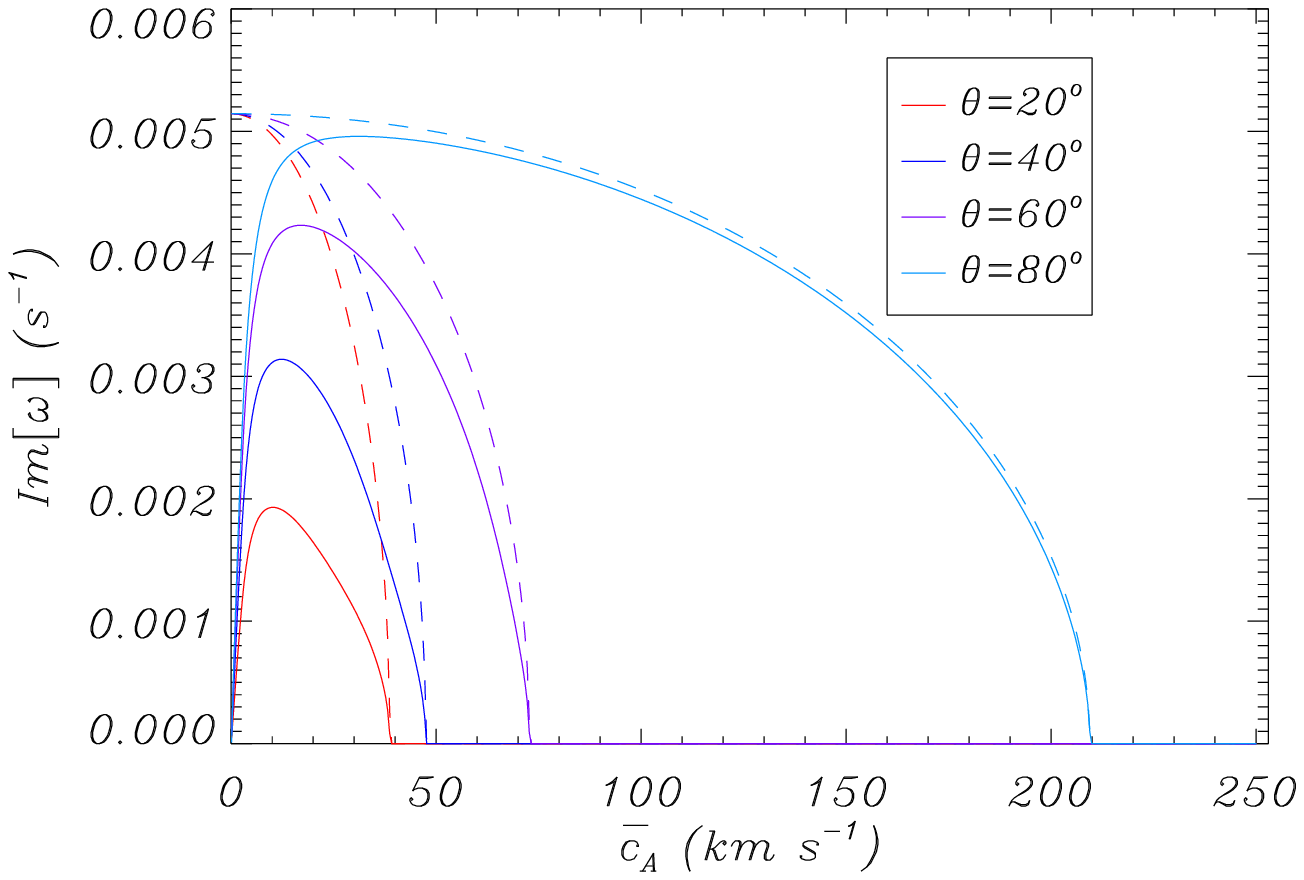} }}
\vspace{-8mm}
  \center{\resizebox{\hsize}{!}{\includegraphics{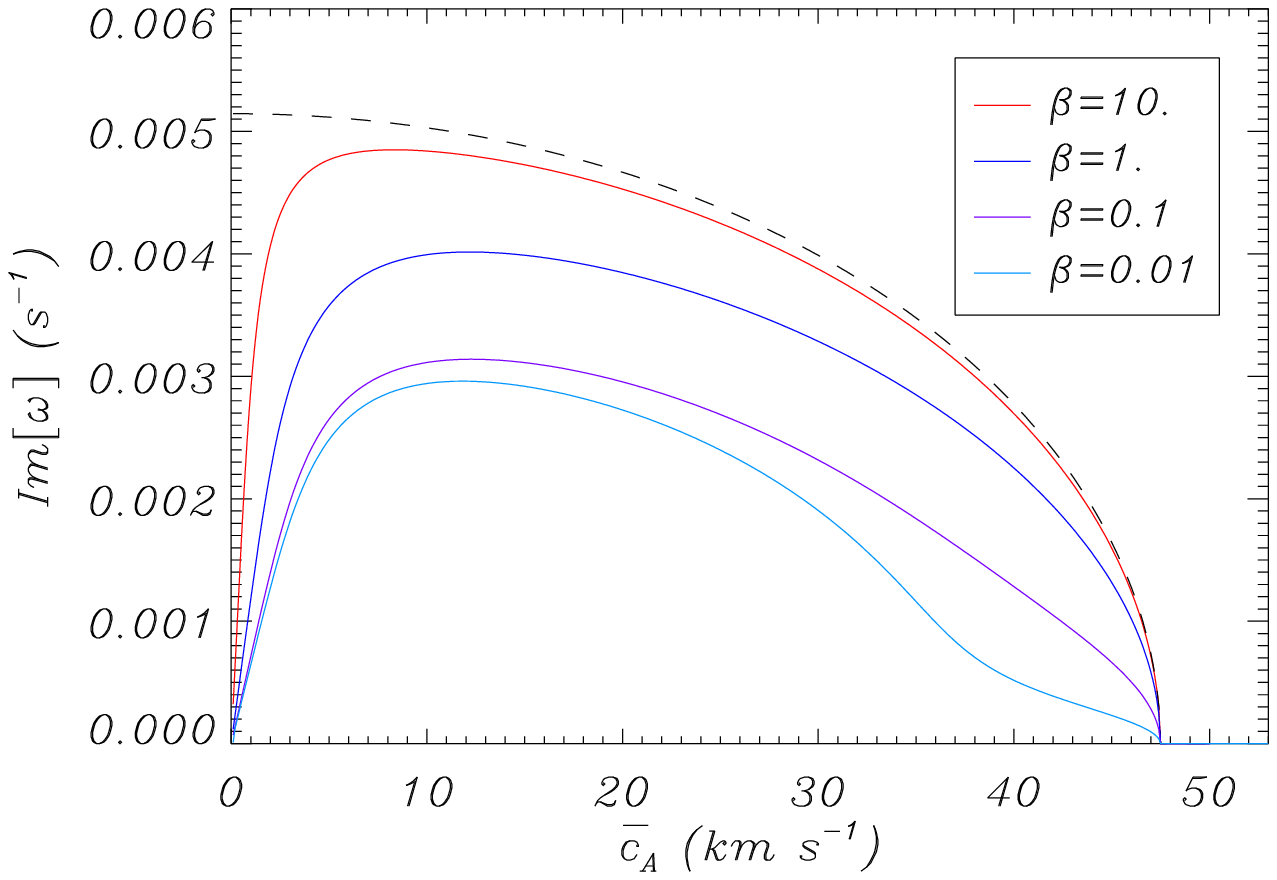} }}
\caption{
Linear growth rate of the RTI for a fully ionized plasma (ideal MHD)
as a function of the Alfv\'en speed $\cv$.
In the upper panel curves
for different values of the propagation angle $\theta$ are shown for a fixed
value of $\beta=0.1$, while in the lower panel curves for different $\beta$ are
plotted for a fixed value of $\theta=40^\mathrm{o}$. In all the panels the values
$\rho_{2}/\rho_{1}=100$, $g=270$ m s$^{-2}$ and $k=10^{-7}$ m$^{-1}$ have been used.
The dashed curves correspond to the incompressible MHD limit in Eq.~(\ref{RTI}).
}
\label{plot_sc}
\end{figure}

After checking the limiting cases, we proceed to solve directly Eq.~(\ref{dr_mhd}).
The linear growth rate is plotted in Fig.~\ref{plot_sc}, compared with the predicted
rate from the classical formula in Eq.~(\ref{RTI}) for the incompressible limit.
The main conclusions from these results are:

\begin{enumerate}

\item
The threshold is not modified by compressibility. This can be easily
demonstrated by noticing that in ideal MHD the frequency of the modes is either
real or pure imaginary \citep{gp04book,gp10book}, so the transition from a
stable to an unstable situation is necessarily at the points in which $\omega=0$
is satisfied. Inserting this condition in Eq.~(\ref{dr_mhd}) we immediately
recover
\begin{equation}
g = \frac{k_x^2 (\rho_1 c_\mathrm{A1}^2 + \rho_2
c_\mathrm{A2}^2)}{(\rho_2-\rho_1) k}, \,\,\,\mathrm{or} \,\,\,
\cv=c_\mathrm{crit}/\mathrm{cos} \theta,
\end{equation}

\noindent
which matches with the stability criteria from Eq.~(\ref{RTI}) and
Eq.~(\ref{stab_crit}). A magnetic field
increase has a stabilizing effect, while increasing the angle between the
equilibrium field and the wavevector has the opposite effect,
as it happens in the incompressible limit.

\item
The incompressible approximation becomes more valid as $\theta$ approaches
$\pi/2$. This is caused by the fact the incompressible limit is recovered when
$\cssq \to \infty$, which implies $\Omega \to \infty$, so terms containing the
gravity in Eq.~(\ref{m_sc}) and Eq.~(\ref{dr_mhd}) are negligible. We can see in
the definition of $m$ and the dispersion relation that increasing the
longitudinal wavenumber has a similar effect, and hence, the incompressible
limit is a better approximation as $k$ is increased.

\item
The linear growth rate for the compressible case is always below the
incompressible limit prediction. As $\beta$ is lowered the linear growth rate is
decreased substantially.

\end{enumerate}

The curves in Fig.~\ref{plot_sc} tend to zero when the magnetic field is very
low. This is caused by the choice of sound speed: since $\beta$ is fixed in
these curves, $\cv \rightarrow 0$ also implies that $c_\mathrm{s} \rightarrow
0$. If the sound speed is prevented from tending to zero as $B_0 \rightarrow 0$
(implying that $\beta$ is no longer held constant and tends to zero too) the
drop dissapears. There is also a real part of the frequency only when the
configuration is stable in this ideal MHD regime, but we focus on the imaginary
part and the instability. Leaky modes may also be considered (i.e. modes that
propagate in the direction across the surface), but these modes do not appear
for the range of parameters selected in these plots.

\subsection{Partially Ionized Plasma}

Now we turn to the general problem with the ambipolar diffusion coefficient
different from zero. Eq.~(\ref{mdc}) is a fourth-order ordinary
differential equation with constant coefficients, whose solutions are a combination
of exponentials $e^{m_k z}$, with $m_k$ one of the four solutions to the indicial
equation
\begin{equation}
C_4 m^4 + C_3 m^3 + C_2 m^2 + C_1 m +C_0 =0.
\label{indicial}
\end{equation}

\noindent
Two of these solutions are close to those in Eq.~(\ref{m_sc}), while the other two
are typically larger and depend strongly on the exact value of $\etaa$.
Eq.~(\ref{indicial}) must be solved in each of the two regions, and then only the
two solutions that imply evanescence away from the discontinuity are kept. The
general expression of the four roots of this fourth order algebraic equation is
massive, so we choose to solve it numerically. Our general solution is then
\begin{equation}
b_x (z)=\left\{\begin{array}{ll}
    A_{1} \, e^{m_{1}^{(1)} z} + A_{2} \, e^{m_{2}^{(1)} z} , & z < 0 , \\
    A_{4} \, e^{m_{3}^{(2)} z} + A_{3} \, e^{m_{4}^{(2)} z}, & z > 0 .
    \end{array} \right.
\label{4ord_sol}
\end{equation}

\noindent
with the $A$ coefficients being arbitrary constants and the subscript of $m$
denoting the ordering of the real part among the set of $m_k$ and the superscript
the region where it applies.

The boundary conditions must be applied to obtain a
dispersion relation, taking into account that $v_z$ must be obtained by integrating
Eq.~(\ref{vz_eq}) after inserting the solution for $b_x$ in Eq.~(\ref{4ord_sol}).
Using the same notation that in \citet{dsb12}, the four boundary conditions are
written in matricial form as
\begin{equation}
\left[ \sum_{j=1,4} B_{ij} b_x^{(3-j)} \right]= 0,
\label{bc_mat}
\end{equation}

\noindent
where the index $i \in [1,4]$ stands for the each boundary condition  in
Eqs.~(\ref{bc}) and $b_x^{(j)}$ is the $j$-th $z$-derivative of $b_x$ ($j=0$
being the function itself without derivatives and $j=-1$ the first integral of
the function). The coefficients in this matrix are given in the Appendix.
Inserting the solutions from Eq.~(\ref{4ord_sol}) in Eq.~(\ref{bc_mat}) we obtain a
system of equations for the $A$-coefficients,
\begin{equation}
\sum_{j=1,4}  (-1)^{h_k} A_k B^{(h_k)}_{ij} \left( m_k^{(h_k)} \right)^{3-j}=0,
\end{equation}

\noindent
where the $m$-coefficients are defined in Equation~(\ref{indicial}), with the
requirement that the exponentials are bounded at $z\rightarrow \pm \infty$. In
this expression $h_k=1$ for $k=1,2$ and $h_k=2$ for $k=3,4$
The dispersion relation of the system is obtained by requiring the determinant
of such system to vanish, namely
\begin{equation}
\left| C_{ik}   \right| = 0,
\label{full_dr}
\end{equation}

\noindent
with the $C$-matrix defined as
\begin{equation}
C_{ik}= (-1)^{h_k} \sum_{j=1,4} B_{ij}^{(h_k)} \left( m_k^{(h_k)} \right)^{3-j}.
\label{dr}
\end{equation}

\begin{figure}[h]
  \center{\resizebox{\hsize}{!}{\includegraphics{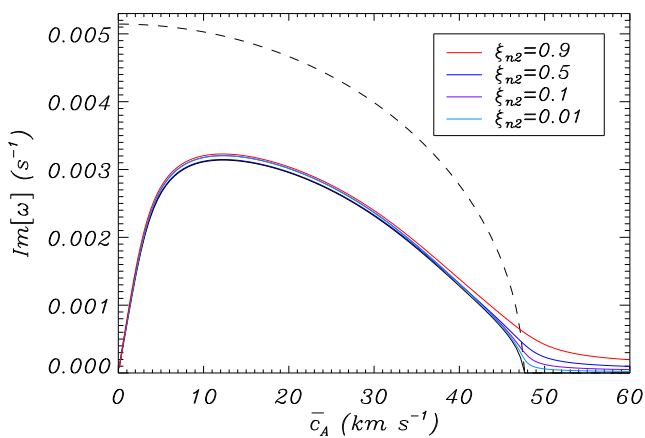} }}
\caption{Linear growth rate of the RTI for a partiallyl ionized plasma (taking
into account ambipolar diffusion)
as a function of the Alfv\'en speed $\cv$.
The dashed line corresponds
to the incompressible limit given in Eq.~(\ref{RTI}) and the black solid line to
the compressible MHD results from Sect.~\ref{MHD_limit_sect}.
The values $\rho_{2}/\rho_{1}=100$, $\beta=0.1$, $\theta=40$, $k=10^{-7}$
m$^{-1}$, $c_\mathrm{crit}=33$ km s$^{-1}$ and $\xi_{n1}=10^{-4}$
have been used.}
\label{wvsck}
\end{figure}

The imaginary part of the solutions to Eq.~(\ref{full_dr}) are plotted in
Fig.~\ref{wvsck}, with the growth rates predicted by the incompressible and
compressible RTI overplotted. The following points must be emphasized:

\begin{enumerate}

\item
A similar plot to Fig.~\ref{wvsck} with higher values of $\theta$ would draw the
collisional plasma results closer to the incompressible MHD limit and modify the
critical speed. Thus, the incompressible limit is a much better approximation as
$\theta$ is increased, as happened in the collisionless plasma.

\item
The instability threshold is no longer the one predicted by Eq.~(\ref{RTI}),
namely $\cv = c_\mathrm{crit}/\mathrm{cos} \theta = 47.4$ km s$^{-1}$ for the
parameters used in the plot. In
fact, {\em the configuration is unstable for all the values of $\xi_n$ and
magnetic field}. Considering the presence of neutrals only with the linear
ambipolar term is enough to render the
configuration with a heavier partially ionized fluid unstable, no matter how
strong the magnetic field is.

\item
For parameters which are classically unstable ($\cv <
c_\mathrm{crit}/\mathrm{cos} \theta$) the linear growth rate is much reduced
with respect to Eq.~(\ref{RTI}) because of the compressibility. Increasing the
ambipolar coefficient raises slightly the growth rate, but the effect of the
ambipolar term is small compared with compressibility in this range.

\item
For parameters which are classically stable ($\cv >
c_\mathrm{crit}/\mathrm{cos} \theta$) the ambipolar diffusion still drives the
instability, but the linear growth rate in this regime is an order of magnitude
smaller than in the classically unstable range.

\item
Close to the stability threshold ($\cv \approx c_\mathrm{crit}/\mathrm{cos}
\theta$) the differences induced by the ambipolar diffusion term are relatively
higher.

\item
In any case, in all the parameter space the growth rate is significantly lower
than the one of an uncoupled neutral gas subject to the hydrodynamic RTI
(Eq.~\ref{RTI} with $B_0=0$, $\rho_{n1}$ and $\rho_{n2}$), which would be
$Im[\omega]=0.0053$ s$^{-1}$ for the parameters in the plot. The collisional
coupling between neutrals and charged particles prevents the neutrals from fully
developing their instability, even for values of $\xi_{n2}$ close to 1.

\item
As the ionization fraction is raised (and thus $\xi_{n2}$ and $\xi_{n1}$ are
lowered) the curve resembles more the MHD limit. In fact, there is a bifurcation
very close to the critical value which can not be clearly seen in the scale of
these plots and its value tends to $c_\mathrm{crit}/\mathrm{cos} \theta$ as the
neutral fraction tends to zero.

\end{enumerate}

Note that with the inclusion of the ambipolar term the frequencies of the modes
are no longer restricted to be either pure real or pure imaginary as in the
ideal MHD limit (collisionless plasma). The solutions plotted in
Fig.~\ref{wvsck} have a real conterpart $Re[\omega]$ not shown in the plot,
which is close to the compressible  MHD results when  $\cv >
c_\mathrm{crit}/\mathrm{cos} \theta$ and much smaller than $Im[\omega]$ when
$\cv < c_\mathrm{crit}/\mathrm{cos} \theta$. We can check that there is no
critical value of $\cv$ for which the system becomes stable: if we require
$\omega \to 0$, the only real  solution is $c_\mathrm{crit}=0$, confirming the
numerical results in Fig.~\ref{wvsck} and the absence of a stable region in the
parameter space.

\begin{figure}[h]
  \center{\resizebox{\hsize}{!}{\includegraphics{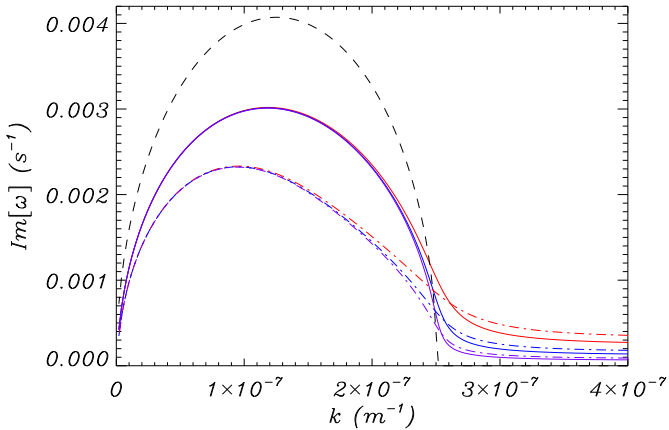} }}
\caption{Linear growth rate of the RTI for a partiallyl ionized plasma
as a function of perturbation wavenumber $k$.
The dashed line corresponds  to the
incompressible limit given in Eq.~(\ref{RTI}).  The values
$\rho_{2}/\rho_{1}=100$, $\theta=40$, $\cv=30$ and $\xi_{n1}=10^{-6}$ have been
used. Solid lines are calculated with $\beta=0.1$ and dot-dashed lines with
$\beta=0.5$, while red lines have a value for the neutral fraction
$\xi_{n2}=0.9$, blue lines $\xi_{n2}=0.5$ and purple lines $\xi_{n2}=0.1$.}
\label{wvsk}
\end{figure}

\begin{figure}[h]
  \center{\resizebox{\hsize}{!}{\includegraphics{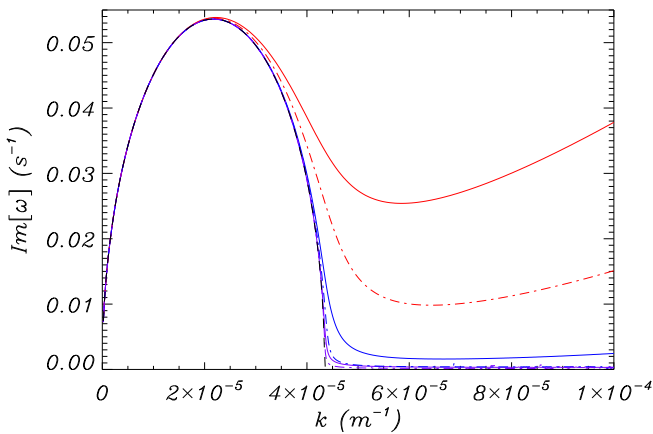} }}
\caption{Same plot as in Fig.~\ref{wvsk} for the values of $\theta=89$ and
$\cv=100$ (near the incompressible limit). Here the values of the beta are
$\beta=0.01$ in the solid lines and $\beta=0.1$ in the dashed lines.}
\label{wvsk_lena}
\end{figure}

Another important parameter is the perturbation wavenumber $k$. So far we have
fixed a value of $k=10^{-7}$ m$^{-1}$, following the typical wavenumbers from
the fast transversal MHD modes of a prominence thread used in previous studies
in prominence seismology and RTI instability in threads \citep{dob02,
tob12,dsb12}, but we can explore further the effects of the initial
perturbation. One important consequence is obtained after scaling the problem:
it can be shown that in the ideal MHD limit the curves can be rescaled using the
variables $\omega /(c k)$ and $\cv/c$ (with $c$ a characteristic
speed, such as Alfv\'en speed in the prominence, for example), but if the
ambipolar diffusion is included, this scaling involves the adimensional quantity
$\etaa k/c$. Hence, increasing the wavenumber perturbation has the direct
effect of increasing the relevance of the ambipolar term. This is expected,
since it is known that the ambipolar diffusion grows as the typical lenght scale
is reduced. On the other hand, we can also plot the linear growth rate as a
function of the perturbation wavenumber (Fig.~\ref{wvsk}). We see the same main
effects: there is no stable regime, the compressibility lowers the growth rate
for $\cv < c_\mathrm{crit}/ \mathrm{cos} \theta$ and the ambipolar diffusion
slightly raises it as $\xi_{n2}$ is increased. It is also interesting to study
this dependence near the incompressible limit with values of $\theta$ close to
90$^o$ (Fig.~\ref{wvsk_lena}); since compressibility is no longer dominant, the
inclusion of the ambipolar term raises the growth rate over the classical RTI
(as reported in \citet{syk13} for the same assumptions that \citet{dsb12} in the
context of local bubble of the solar system).

\begin{figure}[h]
  \center{\resizebox{.95\hsize}{!}{\includegraphics{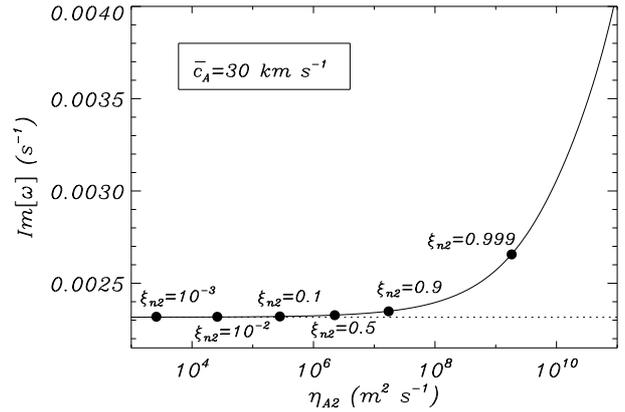} }}
\vspace{-8mm}
  \center{\resizebox{.95\hsize}{!}{\includegraphics{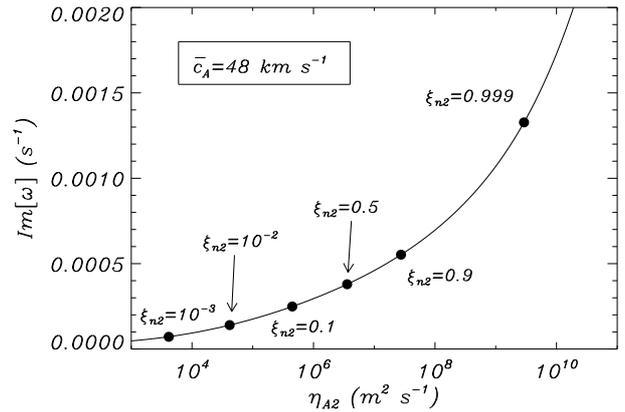} }}
\vspace{-8mm}
  \center{\resizebox{.95\hsize}{!}{\includegraphics{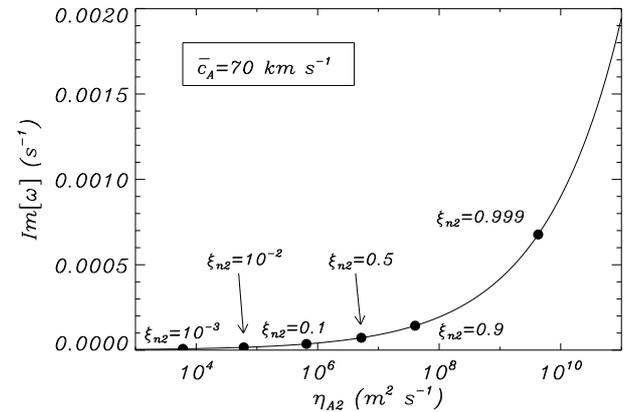} }}
\caption{Linear growth rate of the RTI for a partiallyl ionized plasma
as a function of the ambipolar diffusion coefficient
$\eta_\mathrm{A2}$, with some values
of the neutral fraction $\xi_{n2}$ for which the values of $\eta_\mathrm{A2}$
are obtained also included in the plot as large points.
The values $\rho_{2}/\rho_{1}=100$, $\beta=0.1$, $k=10^{-7}$ m$^{-1}$,
$\theta=40^o$ and
$\xi_{n1}=10^{-6}$ have been used (so $c_\mathrm{crit}=47$ km s$^{-1}$).
The upper panel corresponds to a classically
unstable configuration, the middle panel to a marginally stable configuration
and the lower panel to a classically stable configuration.
The dotted line in the upper panel corresponds to the MHD limit
(in the middle and lower panel the classical limit is zero).
}
\label{wvseta}
\end{figure}

Finally, we can plot the growth rate vs. the ambipolar diffusivity
(Fig.~\ref{wvseta}). As mentioned above, the rate is slightly modified if the
configuration is unstable in the ideal MHD limit ($\cv <
c_\mathrm{crit}/\mathrm{cos} \theta$, upper panel), unless a very
unrealistical high value of the ambipolar diffusivity is assumed.
For configurations that are close to the critical value (middle panel) the
dependence on the ambipolar diffusivity is more important. In the stable range
($\cv > c_\mathrm{crit}/\mathrm{cos} \theta$, lower panel) the linear growth
rate is never zero (so strictly speaking the configuration is unstable),
but the linear growth rate is at least about an order of magnitude lower than
the values in classically unstable regime for typical values of $\etaa$, so in
practice the instability would take an excessively long time to develop.

\section{Full induction equation} \label{sect_fullind}

We have dealt in the previous section with the effect of the induction term and
the ambipolar diffusion term alone in the generalized induction equation for
partially ionized plasmas (Eq.~\ref{ind_etas}). Considering only these terms has
allowed us to solve analytically the linearized equations, but the effect of the
other supposedly smaller terms must be taken into account. In this case,
obtaining analytical solutions can be much more demanding, even in the
linearized problem.

\subsection{Linearized equations}

In order to solve this problem, first we need to eliminate the partial pressures
and the density of each species in the induction equation. This can be done
using the definition of the partial pressure $p_\alpha=n_\alpha k_\mathrm{B}
T_\alpha$, and invoking that the temperature of each species is the same
($T_i=T_e=T_n$). Hence,
\begin{equation}
p=p_i+p_e+p_n=2 p_e + p_n = p_e (2+ \xi_n/\xi_i),
\end{equation}

\noindent
and we obtain
\begin{equation}
p_e=\frac{1-\xi_n}{2-\xi_n} \, p; \,\,\,\,\, p_n=\frac{\xi_n}{1-\xi_n} \, p;
 \,\,\,\,\, \rho_e=\frac{m_e}{m_i} (1-\xi_n) \rho,
\end{equation}

\noindent
so we can operate Eq.~(\ref{G_brag}) to obtain an expression for the pressure
gradients in the battery term,
\begin{equation}
{\bf G}=\frac{\xi_n (1-\xi_n)}{2-\xi_n} \nabla p,
\end{equation}
\begin{equation} \label{battery}
\frac{\varepsilon {\bf G} - \nabla p_e}{e n_e} = \frac{m_e}{e \rho_e}
\frac{1-\xi_n}{2-\xi_n} (\varepsilon \xi_n -1) \nabla p= \frac{m_i}{e} \,
\frac{\varepsilon \xi_n - 1}{2-\xi_n} \, \frac{\nabla p}{\rho}.
\end{equation}

\noindent
Here it is explicit that for a fully ionized plasma ($\xi_n=0$) the ${\bf G}$
combination vanishes and the only contribution to the battery term in that case
is the well-known form of the electron pressure gradient.

Next, we linearize the fluid equations. The only difference with the system in
Eqs.~(\ref{linear_eqs4}) is the linear version of the induction equation.
Regarding the generalized battery term, the curl of Eq.~(\ref{battery}) is

\vspace{-1mm}
\begin{eqnarray}
\nabla \!\!\! &\times& \!\!\! \frac{\varepsilon {\bf G} - \nabla p_e}{e n_e} =
	\frac{m_i}{e} \, \frac{\varepsilon \xi_n - 1}{2-\xi_n}
	\, \nabla \times \frac{\nabla p}{\rho}  \nonumber \\
&=& \frac{m_i}{e} \, \frac{\varepsilon \xi_n - 1}{2-\xi_n} \nabla \frac{1}{\rho}
	\times \nabla p,
\end{eqnarray}

\noindent
and since the equilibrium state has both the density and pressure constant in
each zone, this term is at least of second order in perturbed quantities and
can be neglected in the linear analysis. Taking this into account the linear
version of the induction equation is

\vspace{-1mm}
\begin{eqnarray} \label{lin_gen_ind}
\frac{\partial {\bf b}}{\partial t} \!\!&=&\!\! \nabla \times \left(
	\rule{0mm}{3.5mm}  {\bf v} \times {\bf B}_0 + \eta \nabla^2 {\bf b}+
	\eta_\mathrm{H} \left\{ \left( \nabla \times
	{\bf b} \right) \times{\bf B}_0 \right\} \right. \nonumber \\
&+&\!\! \left. \eta_\mathrm{A} \left\{ \left( \nabla \times
	{\bf b} \right) \times{\bf B}_0 \right\}  \times{\bf B}_0 +
	\eta_G/(\rho_0 \cssq) \nabla p \times {\bf B}_0 \right. \nonumber \\
&-&\!\!	\left. \chi_\mathrm{g1}/\rho_0  \left[ \rho_0{\bf g} \times
	{\bf b} + \rho \, {\bf g} \times {\bf B}_0
	\right] \rule{0mm}{3.5mm} \right),
\end{eqnarray}

\noindent
where $\eta_\mathrm{A}$ is given in Eq.~(\ref{eta_param}) and the other
diffusion coefficients can be expressed also in terms of the ionization fraction
and the equilibrium parameters,

\vspace{-1mm}
\noindent
\begin{eqnarray}
\eta &=& \frac{m_i m_e}{\mu e^2} \frac{1}{\rho_0 (1- \xi_n)} \left[
	\nu_{en} (\xi_n) + \nu_{ei} (\xi_n) \right] - \nonumber \\
   &\,& \,\,\,\,\,\, \frac{m_e^2}{\mu e^2}
	\frac{(\nu_{en} (\xi_n))^2}{\alpha_n (\xi_n)}, \nonumber \\
\eta_\mathrm{H} &=& \frac{m_i}{\mu e} \frac{1}{\rho_0 (1-\xi_n)}
	\left[1 - \frac{\sqrt{2}}{5} \frac{m_e}{m_i} \xi_n \right]
  	, \nonumber \\
\chi_\mathrm{G} &=& \rho_0 \cssq \frac{\xi_n^2}{\alpha_n} \,
	\frac{1-\xi_n}{2-\xi_n}, \nonumber \\
\chi_\mathrm{g1} &=& \rho_0 \frac{\xi_n^2}{\alpha_n} \frac{m_e}{m_i} (1-\xi_n).
\end{eqnarray}

\noindent
Eliminating the perturbed pressure and density we obtain the following set of
partial differential equations for the components of the perturbed velocity and
magnetic field,

\vspace{-1mm}
\begin{eqnarray}
\frac{\partial^2 {\bf v}}{\partial t^2} &=& \frac{1}{\mu \rho_0}
   \left(\nabla \times {\bf b}
   \right) \times {\bf B}_0 - \left(\nabla \cdot {\bf v}\right) {\bf g} +
   \cssq \nabla \left( \nabla \cdot {\bf v} \right), \nonumber \\
\frac{\partial^2 {\bf b}}{\partial t^2} &=& \nabla \times \left(
	\frac{\partial {\bf v}}{\partial t} \times {\bf B}_0 \right) + \eta
	\nabla^2 \frac{\partial {\bf b}}{\partial t} \nonumber \\
   &-& \eta_\mathrm{H} \nabla \times \left\{ \left( \nabla \times
	\frac{\partial {\bf b}}{\partial t} \right) \times{\bf B}_0 \right\}
	\nonumber \\
   &+& \eta_\mathrm{A} \left\{ \left( \nabla \times \frac{\partial {\bf b}}
   	{\partial t}\right) \times{\bf B}_0 \right\}  \times{\bf B}_0
	\nonumber \\
   &+& \chi_\mathrm{G} \nabla \times \left\{ \nabla (\nabla \cdot
   	{\bf v}) \times {\bf B}_0 \right\} \nonumber \\
   &+& \chi_\mathrm{g1} \left[ \left\{ \nabla (\nabla \cdot
	{\bf v} )\right\} \times \left( {\bf g} \times {\bf B}_0 \right) -
	\left( {\bf g} \cdot \nabla \right) \frac{\partial  {\bf b}}
	{\partial t}  \right],
\label{linear_eqs2}
\end{eqnarray}

\noindent
The complexity of these equations is evident, and even third order derivatives
are present in the term coming from ${\bf G}\times{\bf B}$. Finding direct
analytical solutions is still possible in our problem if we notice that all the
coefficients are constant in each region of our model, so we obtain a third
order linear system of six equations, whose solutions are written in terms of
linear combinations of exponential functions.

\subsection{Normal mode analysis}

We again consider the normal mode decomposition with the temporal dependence as
$e^{-i \omega t}$ and the dependence on the directions where the equilibrium
state is uniform as $e^{ik_x x + ik_y y}$. Now we can eliminate $v_x$, $v_y$ and
$b_z$ to obtain the three differential equations, which are given in the
Appendix. We obtain four solutions that go to zero as $z \rightarrow \infty$
and other four that go to zero as $z \rightarrow -\infty$, so the general
solution in each zone would be a linear combination of the four linearly
independent solutions that satisfy the boundary condition as $|z| \rightarrow
\infty$ in each region.

Next we need to derive the boundary conditions appropiate to this problem, and
following the procedure used in the case of ambipolar diffusion alone, we go
directly to Eqs.~\ref{linear_eqs2} and integrate them across the boundary,
obtaining only five relations between the variables

\vspace{-1mm}
\begin{eqnarray} \label{bc1}
\left[ \rule{0mm}{3.5mm} \rho_0 \cssq v_z \right] &=& 0, \nonumber \\
\left[\rule{0mm}{3.5mm} \rho_0 \left\{ i \omega c_\mathrm{A}^2 b_x
    c_\mathrm{s} - i k_x c_\mathrm{s}^2 v_x - i k_y c_\mathrm{s}^2 v_y
    \right. \right. \,\,\,\,&\,& \nonumber \\
   \left.\left. \rule{0mm}{3.5mm} - g v_z + c_\mathrm{s}^2 v_z'
    \right\} \rule{0mm}{3.5mm} \right]&=&0, \nonumber \\
\left[ \rule{0mm}{3.5mm} \eta_\mathrm{A} (i k_x \omega b_z + \omega b_x')
   -\eta \omega b_x'  -i k_x \omega b_y \eta_\mathrm{H}
    \right. \,\,\,\,&\,& \nonumber \\ \left.\rule{0mm}{3.5mm}
    + \chi_\mathrm{G} c_\mathrm{s}^2 (-i k_y^2 v_z + k_x v_x'+k_y v_y' +
    i v_z'')
   \right. \,\,\,\,&\,& \nonumber \\ \left.\rule{0mm}{3.5mm}
   -g \chi_\mathrm{g1} (\omega b_x + k_x v_x +k_y v_y + i v_z')
    - \omega v_z \right] &=&0, \nonumber \\
\left[ \rule{0mm}{3.5mm} i k_x \left( \omega  b_x \eta_\mathrm{H} +k_y
   c_\mathrm{s}^3 \chi_\mathrm{G} v_z \right) -\omega  \eta b_y' -g \omega
   b_y \chi_\mathrm{g1} \right] &=&0, \nonumber \\
\left[ \rule{0mm}{3.5mm} k_x \omega  \eta_\mathrm{A} b_x +i \omega  \eta
   b_z'- i g k_x c_\mathrm{s} \chi_\mathrm{g1} v_z \right.
   \,\,\,\,&\,& \nonumber \\
   \left.\rule{0mm}{3.5mm} +k_x c_\mathrm{s}^3
   \chi_\mathrm{G}  \left(k_x v_x +k_y v_y +i v_z' \right) \right] &=& 0.
\end{eqnarray}

However, these relations are not enough for our problem, which has four
arbitrary constants in each side of the boundary. The divergence-free condition
$\nabla \cdot {\bf b} =0$ and the expressions for the perturbed pressure and
density in Eqs.~\ref{linear_eqs4} only give us linear combinations of the
conditions in Eq.~\ref{bc1}. To obtain the additional relations we need to use
the equation for the diffusion velocity between ions and neutrals (which
introduces the higher-order derivatives in the linearized equations). The
linear version of Eq.~(\ref{eq_w}) is

\vspace{-1mm}
\begin{eqnarray}
{\bf w} &=& \eta_\mathrm{A} \left[ \nabla \times
\frac{\partial {\bf b}}{\partial t}  \right] \times {\bf B}_0 +
\chi_\mathrm{G1} \nabla \left( \nabla \cdot {\bf v} \right) \nonumber \\
&+& \eta_\mathrm{H1}
\nabla \times \frac{\partial {\bf b}}{\partial t} - \chi_\mathrm{g2} (\nabla
\cdot {\bf v}) \, {\bf g} =0,
\end{eqnarray}

\noindent
with the coefficients defined as
\begin{equation}
\eta_\mathrm{H1} = \frac{\xi_n \nu_{ne} \rho_0}{n_e e \mu \alpha_n}, \,\,\,
\chi_\mathrm{G1} = \frac{\rho_0}{\xi_n} \chi_\mathrm{G}, \,\,\,
\chi_\mathrm{g2} = \frac{1}{\xi_n} \chi_\mathrm{g1}.
\end{equation}

\noindent
Using this equation, we obtain the remaining three jump relations for our
system, namely

\vspace{-1mm}
\begin{eqnarray} \label{bc2}
\left[ \rule{0mm}{3.5mm} \alpha_n \left( \omega b_y \eta_\mathrm{H1}
   -i k_x c_\mathrm{s} \chi_\mathrm{G1} v_z \right) \right] &=& 0, \nonumber \\
\left[\rule{0mm}{3.5mm} \alpha_n \left( -i \omega b_x \eta_\mathrm{H1}-
   i k_y c_\mathrm{s} \chi_\mathrm{G1} v_z  \right)\right]&=&0, \nonumber \\
\left[ \rule{0mm}{3.5mm}  \alpha_n \left( i \omega  \eta_\mathrm{A} b_x
   - g c_\mathrm{s} \chi_\mathrm{g2} v_z -i k_x c_\mathrm{s}
   \chi_\mathrm{G1} v_x\right. \right. \,\,\,&\,& \nonumber \\
   \left. \rule{0mm}{3.5mm} \left.- i k_y c_\mathrm{s} \chi_\mathrm{G1} v_y
   +c_\mathrm{s} \chi_\mathrm{G1} v_z'
   \right) \right] &=& 0.
\end{eqnarray}

\noindent
which provides us with the remaining conditions to solve the linear problem.
Notice that we recover easily the case with ambipolar diffusion alone, since
the first two equations vanish in this case and the last one becomes equivalent
to the last jump condition in Eq.~(\ref{bc2_vs}).

\subsection{Numerical solutions}

The solution of the problem must be computed in the following form: first of all,
the differential equations in Eq.~\ref{dif_sist2} for $b_x$, $b_y$ and $v_z$ must
be solved in each zone by obtaining the set of solutions for $\lambda$ from
Eq.~\ref{indicial_sist} (and the relation between the constants of each variable),
and then discarding the solutions that not vanish as $|z| \rightarrow 0$. This
leaves us with solutions with four arbitrary constants in each zone. Finally, the
jump conditions in Eqs.~(\ref{bc1}) and (\ref{bc2}) must be satisfied, which can
only be achieved is the determinant of the system of this eight equation vanishes.
This provides us with a dispersion relation for computing the frequencies of our
system, and thus, studying the stability of the system by checking if their
imaginary parts are either positive (unstable modes) or negative (stable modes).

The procedure described in the previous paragraph does not give simple
analytical solutions, so we use it to find numerical solutions to the system. It
is important to remark that these solutions are not obtained from partial
differential equations, but from an algebraic system of equations. There is a
huge range of parameters that can be explored, but here we concentrate on the
situations of physical interest for the RTI instability, namely, when all the
diffusive terms are much lower than the induction term. All the $\eta$ and
$\chi$ coefficients in these diffusive terms depend on the ionization fraction
$\xi_n$ due to their dependence on the densities, collisions frequencies and
temperatures.

\begin{figure}[h]
  \center{\resizebox{\hsize}{!}{\includegraphics{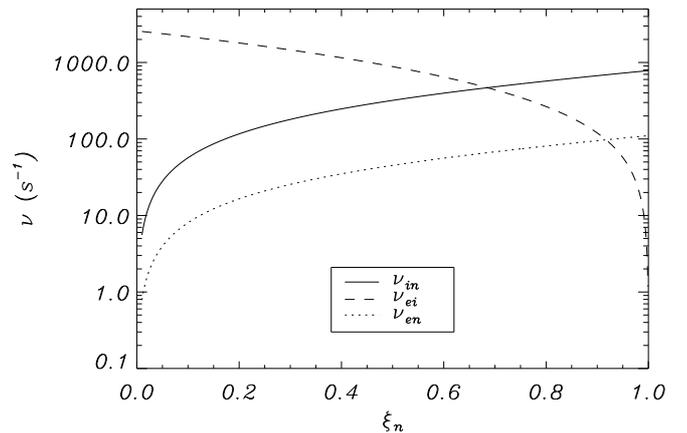} }}
\caption{Collisional frequencies as a function of the ionization fraction for
the values of the density, pressure and magnetic field given in the text.}
\label{freq_num}
\end{figure}

First of all, we study the numerical values and dependence of the different
collision frequencies, since we also need the collisional frequencies of
electrons with other species, which are \citep{sob09b,Braginskii}
\begin{equation}
\nu_{en} = \frac{\rho_n}{m_\mathrm{n}} \sqrt{\frac{16 k_\mathrm{B} T}
   {\pi  m_\mathrm{n}}} \, \sigma_\mathrm{en}, \,\,\,\,\,
\nu_{ei} = \frac{n_e}{T^{3/2}} \Lambda  h_\mathrm{ei},
\end{equation}

\noindent
where $\sigma_\mathrm{en} \approx 10^{-19}$ m$^2$, $h_\mathrm{ei}  \approx 3.7
\times 10^{-6}$ s$^{-1}$ m$^{-3}$ K$^{3/2}$ and $\Lambda$ is the Coulomb
logarithm. We plot these frequencies in terms of the ionization fraction in
Fig.~\ref{freq_num} for values of the parameters typical in prominences, namely
$\rho_0=10^{-10}$ kg m$^{-3}$,
$p_0=0.135$ Pa 
and $B_0=10$ G, so $c_\mathrm{s}=15$ km s$^{-1}$
and $c_\mathrm{A}=89$ km s$^{-1}$. For fully ionized plasmas the neutral collision
frequencies vanish, as expected, but as the ionization fraction is increased it
becomes comparable and even larger than the collision frequency between ions and
electrons. Note that the assumption in \citet{sdgb12,dsb12} of neglecting the
electron collisions might not be appropiate for these values of the plasma
parameters (specially for low values of $\xi_n$).

\begin{figure}[h]
  \center{\resizebox{\hsize}{!}{\includegraphics{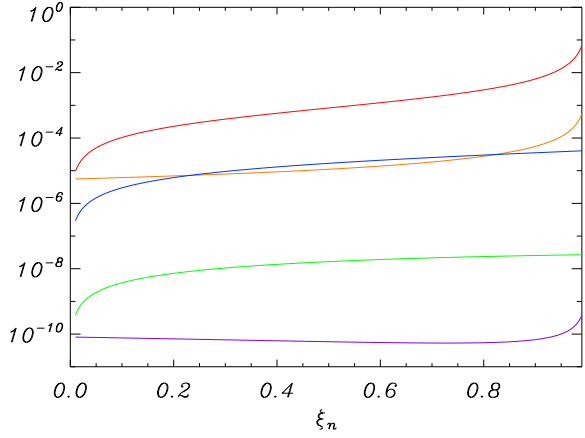} }}
\caption{Numerical values of the terms in the induction equation in
Eq.~(\ref{linear_eqs2}) relative to the ideal MHD term, with the values stated in
the text. The red line corresponds to the ambipolar term, the blue line to the
Hall term, the orange line to the perpendicular battery term (${\bf G} \times {\bf
B}$), the green line to the gravity terms and the purple line to the Ohmic term.}
\label{eta_num}
\end{figure}

Next we study the different terms in the linearized induction equation
(Eq.~\ref{linear_eqs2}). The values of each term are represented in
Fig~\ref{eta_num} for typical values of $k_x=k_y=10^{-7}$ m$^{-1}$,  $g=270$ m
s$^{-1}$ and $Im[\omega] \approx 0.007$ s$^{-1}$, normalized to the magnitude of
the ideal MHD induction term $\nabla \times ({\bf v} \times {\bf B}_0)$. The
dominant term under these plasma conditions is the ambipolar term, which was
studied independently in the previous section. Then the Hall and perpendicular
battery terms are typically about one order of magnitude smaller than the
ambipolar, and finally the gravity and ohmic diffusion terms are much lower.
Note that for a fully ionized plasma $\xi_n \to 0$ the ambipolar, battery and
gravity terms tend to zero, but the hall and ohmic terms are still present. In
addition, the battery term neglected in the linearization (Eq.~\ref{battery})
would also be present for a fully ionized plasma.

Next we study the solutions of the indicial equation  (Eq.~\ref{indicial_sist}).
There are four solutions that are very close to the ones in
Eq.~(\ref{indicial}), and four new ones that are related to the other diffusion
coefficients and are about four order of magnitude larger (and thus describe
only diffusive effects very near the boundary and are negligible far from it).
However, these diffusive solutions are troublesome from the computational point
of view, since they introduce large coefficients in the boundary conditions that
must be computed with great accuracy.

\begin{figure}[h]
  \center{\resizebox{\hsize}{!}{\includegraphics{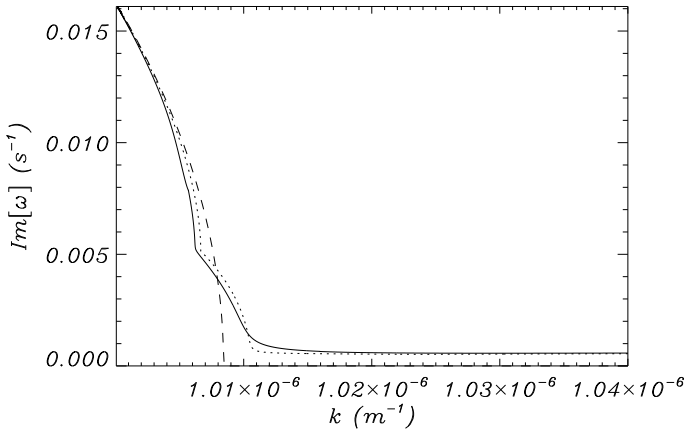} }}
\caption{Linear growth rate of the RTI as a function of the wavenumber.
The values chosen for the plot are $\rho_2=10^{-10}$ kg m$^{-3}$,
$\rho_2/\rho_1=100$,
$\theta=85^o$ m$^{-1}$, $B_0=10$ G, $c_\mathrm{s2}$=15 km s$^{-1}$, $\xi_2=0.5$
and $\xi_1=0.1$. The dashed line is the incompressible MHD limit from
Eq.~(\ref{RTI}), the dotted line corresponds to the PI 1-fluid model with only
the ambipolar term (section~\ref{sect_ambip}) and the solid line to the PI
1-fluid model with all the terms in the induction equation.
}
\label{wvskx}
\end{figure}

Finally we can obtain the frequency of the modes of the system. We concentrate
on the relevant modes to the stability analysis. The imaginary part of the
frequency is plotted near the instability threshold in Fig.~\ref{wvskx} for a
typical set of parameters in prominence thread oscillations. The differences
between the ambipolar result and the full equations are small, but one
interesting difference is that in the MHD stable regime (which corresponds to
$k_x>1.3 \cdot 10^{-7}$ m${-1}$ for these parameters) the inclusion of the rest
of the term in the induction equation raises the linear growth rate slightly,
while in the MHD unstable regime ($k_x<1.3 \cdot 10^{-7}$ m${-1}$) it lowers it
slightly, but the corrections are small compared with the computed values for
the ambipolar case. Hence, as expected from Fig.~\ref{eta_num} the rest of the
terms offer just slightly corrections to the results for the ambipolar case
described in Sect.~\ref{sect_ambip}, at least for the physical and plasma
parameters in prominences.

It is very interesting to notice that the PI effects do not modify
qualitatively the linear growth rate in the region of classical stability.
This is in contradiction with the result in \citet{dsb12}, who reported in
their Section~5 that the linear growth rate was lowered by an order of
magnitude. However, in that calculation a very low value of the equilibrium
density was chosen to compute $\nu_{in}$, so it would correspond to the case in
which the ambipolar coefficient is chosen to be larger than the value obtained
in this work. We can see in Fig.~\ref{wvskx} that the effects of PI do not lower
that drastically the linear growth rate for this set of parameters. We obtain
from Fig.~\ref{wvskx} a typical RTI timescale of about 100 s, similar to the
lifetime of prominence threads \citep{lhvkpgsk10,mkbsa10,lin2011}.

\section{Discussion and conclusions}

We have studied the effects of considering a partially ionized plasma in the MHD
Rayleigh-Taylor instability in a contact surface where a heavier plasma sits on
top of a lighter one. We have simplified considerably the problem by assuming
that the equilibrium variables are uniform in each region, which is only valid
if the vertical scales of the perturbations are much smaller than the
gravitational scale height. In fact, in the initial stages of the instability
the solution is confined to the boundary, so this approximation is useful.
However, in later stages the gravity stratification may become important, but
then the differential equations may become too hard to be solved analytically,
and the problem is better posed in terms of numerical studies, which would also
allow to characterize the non-linear phases of the instability.

Including PI effects in the MHD equations can be done in several ways. In
\citet{dsb12} a two-fluid model was considered, with the collisions between ions
and neutrals only deemed important. Here we have taken a different approach by
using all the collision frequencies between the species, but combining the fluid
equations for each species into 1-fluid equations \citep[following][for
example]{Braginskii}. These PI effects appear then in the form of a generalized
Ohm's law (Eq.~\ref{ohmlaw}) and induction equation (Eq.~\ref{eq_induction}),
with the corresponding terms in the energy equation (which are second order
effects in the linear analysis). We follow the standard procedure of deriving a
relation for the diffusion velocity between ions and neutrals from the equation
of motion for electrons (with the electron inertial terms neglected). However,
in contrast with previous deductions we have kept all the terms, obtaining the
well known expressions for the ohmic, ambipolar, Hall and battery diffusion
terms, but also the ${\bf G} \times {\bf B}$ (similar to the Hall term with
diamagnetic currents) and the gravity term. This gravity effect has been
normally overlooked because it comes from neglecting the electron gravity force
in front of the ion gravity force on the combined momentum equation for ions and
electrons, but we have proved that this term survives as the equation for the
diffusion velocity is obtained. Under prominence thread circumstances, this term
is nevertheless small, but can be still larger than the ohmic diffusion, and
might also be relevant in other contexts.

It has been previously assessed that the most important term in the generalized
Ohm's law is the ambipolar diffusion term \citep{kc12}. We first study the
modifications that this term implies in the linear regime. An ordinary
differential equation is derived with constant coefficients because of the
uniform plasma assumption in each zone, so a solution close to the ideal-MHD is
found, with another related to the ambipolar coefficient. The ordinary MHD jump
relations are not enough, so following \citet{chandrasekhar} we derive our
boundary relations directly from the differential equations, obtaining new
conditions to add to the continuity of total pressure are perpendicular
velocity. Finally the modes of the system are obtained, with the MHD-limit
recovered when $\eta_\mathrm{A} \rightarrow 0$. The main conclusions are that
the configuration is always unstable regardless of the values of the parameters,
but in the region of the parameter space where there was classical stability the
linear growth rate is very small, while in the classically stable region the
ambipolar slightly raises the linear growth rate compared with the compressible
MHD limit. These results support the conclusions in \citet{dsb12} and
qualitatively both descriptions agree despite considering different assumptions
on the fluid equations.  Notice however that in the 2-fluid description
\citet{dsb12} considered that the collision frequency of both media were simply
related by $\rho_2/\rho_1$ and used a high value of equilibrium density, while
here we have  derived a relation between the ambipolar diffusion coefficients in
both regions considering all the dependence of the equilibrium parameters on
$\xi_n$ in  both regions.

Next we consider the full induction equation, checking first the relevance of the
different terms in the linear analysis. It is found that the battery and gravity terms
do not give any direct contribution in the linear regime in a uniform equilibrium
medium, but their Hall counterparts still appear. The other terms are orders of
magnitude smaller than the ambipolar term. The problem is solved in a similar way,
with extra solutions to the indicial equation because of these dissipative terms. The
numerical analysis of the solutions confirms that for typical values of prominence
threads they only induce small corrections to the results of the ambipolar case.

\begin{figure}[h]
  \center{\resizebox{\hsize}{!}{\includegraphics{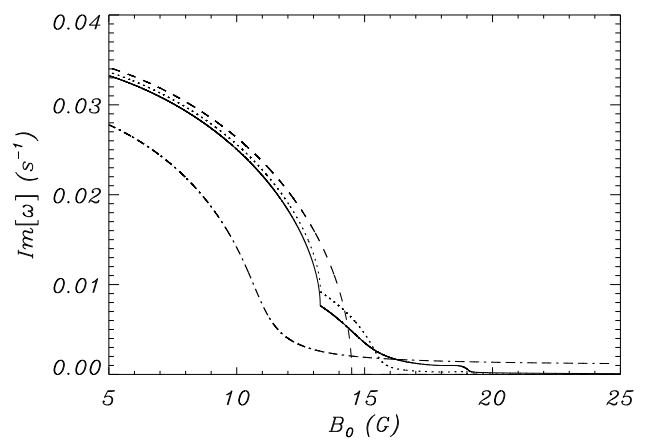} }}
\caption{Linear growth rate vs. equilibrium magnetic field for
$\rho_2/\rho_1=100$, $g=270$ m s$^{-2}$, $\theta=87^o$,
$k=5\cdot 10^{-6}$ m$^{-1}$, $\xi_{n1}=10^{-4}$, $\rho_2=10^{-10}$ kg m$^{-3}$.
The dotted line is the ambipolar case with $\xi_{n2}=0.5$, the solid line with
$\xi_{n2}=0.05$ and the dot-dashed line to $\xi_{n2}=0.5$, but with a value for
the ambipolar diffusion coefficient 1000 times larger than the theoretical value
used in the rest of the computations, while the dashed line
corresponds to the incompressible MHD limit (Eq.~\ref{RTI}).}
\label{wvsb0}
\end{figure}

A direct application of the results of this paper concerns solar prominence
threads. It is widely assumed that chromospheric material sits on top of a less
dense coronal plasma, either in a static equilibrium or dynamical
configurations. The RTI has been studied numerically in such configurations
\citep{hisb11,hbis12}, so it is interesting to test the differences that PI
effects produce, specially taking into account that the material that forms the
prominence is expected to be partially ionized (despite the ionization fraction
has not been directly measured so far). A plot of the linear growth rate for
different values of the equilibrium field is displayed in Fig.~\ref{wvsb0}. The
effects described in our analysis can be summarized as:

\begin{itemize}
\item
There is no critical value, the configuration is always unstable to the RTI
instability because of the presence of neutrals.
\item
On the region of classical stability, the PI terms give a small linear growth
rate, so the time-scale of the instability is much larger than the typical
lifetime of the threads.
\item
On the region of classical instability, the PI affects also the growth rate,
but this rate is still very high (despite a stabilizing effect of the
compressibility), so the RTI is very efficient and can disrupt
the threads.
\item
For typical prominence plasma parameters, the PI effects are small, since the
ambipolar term is much smaller than the MHD induction term (Fig.~\ref{eta_num})
and the perturbation is nearly incompressible \citep{tob12}.
However, if the term is larger than the theoretical values the effect becomes
more pronounced (see the plot for a high $\eta_\mathrm{A}$ value in
Fig.~\ref{wvsb0}).
\item The leading ambipolar term becomes important on small scales
    (for the typical prominence parameters whose scales are expected to
    be in the range of 100 km and below). Current observational
    facilities are almost at this limit (for example the Japanese HINODE
    mission, or Sunrise/IMaX instrument) and the new generation of
    telescopes (such as ATST or EST) are aimed to provide information on
    such scales. Thus, we are about to be able to observe the spatial
    range where the PI effects in prominences might be directly
    observed). 
\item
Including other additional PI terms beyond the leading ambipolar term only give
small numerical changes (mainly near the classical critical value) at the
price of a much harder analytical and computational effort.
\end{itemize}

These conclusions need to be tested in several ways. First of all, the analysis
carried out in this work is only valid in the linear regime, so once the
instability is triggered on, non-linearities may become important, and it is
well-known from MHD simulations that once the stability is well developed
secondary Kelvin-Hemholtz instabilities appear (seen as eddies in the
simulations), so the linear growth rate is lowered and the drops formed reach a
terminal velocity; all these processes are not present in the linear analysis.
Moreover, the battery term contribution can be neglected in the linear analysis,
but helps to raise currents in the non-linear regime. More crucially, no new
effects are present in the linearized energy equation.

Another neglected effect that might be important is the presence of a density
stratification (mainly due to gravity), despite having typical lenghscales much
larger than the thread thickness. Some studies point out that these
stratification effects have a stabilizing contribution on the RTI
\citep{ljtc09}. However, considering a non-uniform plasma in each region
complicates substantially the analysis (specially the differential equations,
which no longer have constant coefficients), and might change the relevance of
some terms (such as the battery term, which would have a linear contribution).
In this case, the problem is better posed to numerical solutions, specially
considering that other effects might also be important, such as the curvature of
the field lines forming the dip that sustains the condensation which constitutes
the thread.

Numerical simulations are underway to study the complex effects of PI in the
instability and the non-linear regime \cite{kdcd13}. The calculations in this
work offer a guide to test the results, at least in the first stages of the
instability.

\begin{acknowledgements}
The authors acknowledge the
financial support by the Spanish Ministry of Science through project
AYA2010-18029. This work contributes to the deliverables identified
in FP7 European Research Council grant agreement 277829,
``Magnetic connectivity through the Solar Partially Ionized
Atmosphere'', whose PI is E. Khomenko (Milestone 3 and
contribution toward Milestones 1).
\end{acknowledgements}

\begin{appendix}

\section{Coefficients on the linear system of equations}

Here we present the coefficients that appear in the linearized equations
for both the ambipolar diffusion section and the general case. If only ambipolar
diffusion is considered in the induction equation (Sect~\ref{sect_ambip}), then
the boundary conditions can be written in terms of $b_x$, as is shown in
Eq.~\ref{bc_mat} 
\vspace{-1mm}
\noindent
\begin{eqnarray}
B_{11} \!&=&\! 0, \nonumber \\
B_{12} \!&=&\! \left\{ i (\omega + i k_x^2 \etaa)[-k^2 \casq (\omega^2 - k_x^2
\cssq) \right. \nonumber \\
&+& \left. \omega (\omega^2 - k^2 \cssq) (\omega + k^2 \etaa) ]\right\}/\left\{
(\omega^2-k_x^2 \cssq) (\omega^2 \right. \nonumber \\
&-& \left. k_x^2 \casq + i \omega \etaa k_x^2)\right\},
\nonumber \\
B_{13} \!&=&\! 0, \nonumber \\
B_{14} \!&=&\! \etaa \frac{-k_x^2 \casq(\omega^2-k_x^2 \cssq)+ \omega (\omega^2
- k^2 \cssq)(\omega + i k_x^2 \etaa)}{(\omega^2 - k_x^2 \cssq)(\omega^2 - k_x^2
\casq + i \omega \etaa k_x^2)}, \nonumber \\
B_{21} \!&=&\! \frac{\omega^2 \cssq \etaa}{\omega^2-k_x^2 \cssq},  \nonumber \\
B_{22} \!&=&\! g \etaa \frac{k_x^2 \casq (\omega^2-k_x^2 \cssq) - \omega
	(\omega^2 - k^2 \cssq) (\omega + i k_x^2 \etaa)}
	{(\omega^2-k_x^2 \cssq)(\omega^2 - k_x^2
	\casq + i \omega \etaa k_x^2)} , \nonumber \\
B_{23} \!&=&\! i \omega \frac{\omega^2 (\casq+\cssq) + i \omega k^2
        \cssq \etaa -k_x^2
	\casq \cssq}{\omega^2-k_x^2 \cssq}, \nonumber \\
B_{24} \!&=&\! - g (i\omega - k_x^2 \etaa) \times \nonumber \\
	&\,& \frac{- k^2 \casq (\omega^2 - k_x^2 \cssq) + \omega (\omega^2 -
	k^2 \cssq) (\omega + i k^2 \etaa)}{(\omega^2-k_x^2 \cssq)(\omega^2 - k_x^2
	\casq + i \omega \etaa k_x^2)}, \nonumber \\
B_{31} \!&=&\! 0, \nonumber \\
B_{32} \!&=&\! \frac{k_y^2 \omega \cssq \etaa}{(\omega^2 - k_x^2 \cssq)
	(\omega^2 - k_x^2 \casq + i \omega \etaa k_x^2)}, \nonumber \\
B_{33} \!&=&\! 0, \nonumber \\
B_{34} \!&=&\! \frac{i k^2 \casq (\omega^2 - k_x^2 \cssq) + \omega (\omega^2 -
	k^2 \cssq) (-i \omega + k^2 \etaa)}{(\omega^2 - k_x^2 \cssq)
	(\omega^2 - k_x^2 \casq + i \omega \etaa k_x^2)}, \nonumber \\
B_{41} \!&=&\! 0, \nonumber \\
B_{42} \!&=&\! 0, \nonumber \\
B_{43} \!&=&\! \etaa, \nonumber \\
B_{44} \!&=&\! 0,
\end{eqnarray}
\noindent
These coefficients are then used in Eq.~\ref{dr} to compute the growth
rates.

Next we present the coefficients that appear in the linearized equations
for the general case, when all the terms in the induction equation
are considered (Sect.~\ref{sect_fullind}). The method is
similar to the one applied in the ambipolar case, but the complexity of the
problem prevents us to write the boundary conditions in terms of a single
magnitude, as it was done in the ambipolar section using $b_x$. First of all,
from the linearized equation of motion we express $v_x$, $v_y$ and $b_z$ as

Eq.~\ref{lin_gen_ind}
\vspace{-1mm}
\noindent
\begin{eqnarray}
v_x &=& \frac{k_x c_\mathrm{s} \left(k_y c_\mathrm{A}^2 (k_x b_y -k_y b_x)+
   i \omega c_\mathrm{s} v_z' \right)}{\omega^3- \omega c_\mathrm{s}^2
   \left(k_x^2+k_y^2\right)},   \nonumber \\
v_y &=& \frac{c_\mathrm{A}^2 \left(k_x^2 c_\mathrm{s}^2-\omega^2\right) (k_y
   b_x-k_x b_y)+ i k_y \omega  c_\mathrm{s}^3 v_z'} {\omega^3 c_\mathrm{s}-
   \omega  c_\mathrm{s}^3 \left(k_x^2+k_y^2\right)}, \nonumber \\
b_z &=& \left[ i k_x^2 c_\mathrm{A}^2 c_\mathrm{s}^2 b_x' - i \omega^2
   c_\mathrm{A}^2 b_x' +i g k_y^2 c_\mathrm{A}^2 b_x +i k_x k_y c_\mathrm{A}^2
   c_\mathrm{s}^2 b_y' \right. \nonumber \\
 &-& \left. i g k_x k_y c_\mathrm{A}^2 b_y +g \omega  c_\mathrm{s}
   v_z' +k_x^2 \omega  c_\mathrm{s}^3 v_z +k_y^2 \omega  c_\mathrm{s}^3
   v_z \right. \nonumber \\
 &-& \left. \omega  c_\mathrm{s}^3 v_z'' -\omega^3 c_\mathrm{s}
   v_z \right]/\left[k_x c_\mathrm{A}^2 \left(c_\mathrm{s}^2 \left(k_x^2+k_y^2
   \right)-\omega^2\right)\right],
\end{eqnarray}

\noindent
Next, we can write each of the three components of the induction equation as
\begin{equation} \label{dif_sist2}
\sum_{j=1}^4 \alpha_{ij} b_x^{(j-1)} +\sum_{j=1}^4 \beta_{ij} b_y^{(j-1)} +
\sum_{j=1}^5 \gamma_{ij} v_z^{(j-1)} =0,
\end{equation}

\noindent
where $i=1,2,3$ is the component of the induction equation and $b_x^{(j)}$
stands for the $j^\mathrm{th}$-derivative of $b_x$, for example. The
coefficients in this equations are

\vspace{-1mm}
\noindent
\begin{eqnarray}
\alpha_{11} &=& c_\mathrm{A}^2 c_\mathrm{s} \left[i \omega
   \left\{ c_\mathrm{s}^2 \left(k_x^2+k_y^2\right) \left(-k_y^2
   \eta _\mathrm{A} +\left(k_x^2+k_y^2\right) \eta+ i
   \omega \right) \right. \right. \nonumber \\
  &+& \left. \left. k_y^2 \omega^2 \eta_\mathrm{A} + i g k_y^3 \eta_\mathrm{H}
   -\omega^2 \left(k_x^2+k_y^2\right) \eta \right\} \right. \nonumber \\
  &+& \left. k_y^2 c_\mathrm{A}^2
   \left(-\omega^2+c_\mathrm{s}^2 \left(k_x^2-i k_y^2 \omega  \chi_\mathrm{G}
   \right)\right)+\omega ^4\right], \nonumber \\
\alpha_{12} &=& -i \omega  c_\mathrm{A}^2 c_\mathrm{s} \left[ g
   \chi_\mathrm{g1} \left(k_y^2 c_\mathrm{A}^2+c_\mathrm{s}^2 \left(k_x^2+
   k_y^2\right)-\omega^2\right) \right. \nonumber \\
  &+& \left. g k_y^2 \eta_\mathrm{A} +i k_y
   \eta_\mathrm{H} \left(\omega^2-k_x^2 c_\mathrm{s}^2\right)\right],
   \nonumber \\
\alpha_{13} &=& i \omega  c_\mathrm{A}^2 c_\mathrm{A} \left[c_\mathrm{A}^2
   \left(k_y^2 \left(c_\mathrm{A}^2 \chi_\mathrm{G}+\eta_\mathrm{A} \right)
   -\left(k_x^2+k_y^2\right) \eta \right)+\omega^2 \eta \right], \nonumber \\
\alpha_{14} &=& 0, \nonumber \\
\alpha_{21} &=& k_x k_y c_\mathrm{A}^2 c_\mathrm{s} \left[c_\mathrm{A}^2
   \left(\omega^2-c_\mathrm{s}^2 \left(k_x^2-i k_y^2 \omega  \chi_\mathrm{G}
   \right)\right) \right. \nonumber \\
  &+& \left. i \omega  \eta_\mathrm{A} \left(c_\mathrm{s}^2
   \left(k_x^2+k_y^2\right)-\omega^2\right)+g l \omega  \eta_\mathrm{H} \right],
   \nonumber \\
\alpha_{22} &=&   -k_x k_y^2 \omega c_\mathrm{A}^2 c_\mathrm{s}^3
   \eta_\mathrm{H}, \nonumber \\
\alpha_{23} &=& \alpha_{24}=0, \nonumber \\
\alpha_{31} &=& \omega  c_\mathrm{A}^2 c_\mathrm{s} \left[g k_x^2
   \chi_\mathrm{g1} \left(k_y^2 c_\mathrm{A}^2+c_\mathrm{s}^2 \left(k_x^2+
   k_y^2 \right)-\omega^2\right) \right. \nonumber \\
  &+& \left. g k_x^2 k_y^2 \eta_\mathrm{A}
   +k_y \left(-g k_y \left(k_x^2+k_y^2 \right) \eta \right. \right. \nonumber \\
  &-& \left. \left. i \left(k_x^2
   \eta_\mathrm{H} \left(c_\mathrm{s}^2 \left(k_x^2+k_y^2\right)-\omega^2
   \right)+g k_y \omega \right)\right)\right], \nonumber  \\
\alpha_{32} &=&   \omega  c_\mathrm{A}^2 c_\mathrm{s} \left[ \omega^2
   \left(\left(k_x^2+k_y^2\right) \eta +i \omega \right) \right. \nonumber \\
  &-& \left. k_x^2 c_\mathrm{s}^2
   \left(k_y^2 \left(c_\mathrm{A}^2 \chi_\mathrm{G} +\eta_\mathrm{A} +\eta
   \right)+k_x^2 \eta+i \omega \right)\right], \nonumber \\
\alpha_{33} &=& g k_y^2 \omega  c_\mathrm{A}^2 c_\mathrm{s} \eta, \nonumber \\
\alpha_{34} &=& \omega  c_\mathrm{A}^2 c_\mathrm{s} \eta \left(k_x^2
   c_\mathrm{s}^2-\omega^2\right), \nonumber
\end{eqnarray}
\vspace{-1mm}
\noindent
\begin{eqnarray}
\beta_{11} &=& k_x k_y c_\mathrm{A}^2 c_\mathrm{s} \left[ c_\mathrm{A}^2
   \left(\omega^2-c_\mathrm{s}^2 \left(k_x^2-i k_y^2 \omega \chi_\mathrm{G}
   \right)\right) \right. \nonumber \\
  &+& \left. i \omega \eta_\mathrm{A} \left(c_\mathrm{s}^2
   \left(k_x^2+k_y^2\right)-\omega^2 \right)+g k_y \omega  \eta_\mathrm{H}
   \right], \nonumber \\
\beta_{12} &=& k \omega  c_\mathrm{A}^2 c_\mathrm{s} \left(\eta_\mathrm{H}
   \left(k_x^2 c_\mathrm{s}^2-\omega^2 \right)+i g k_y \left(c_\mathrm{A}^2
   \chi_\mathrm{g1}+ \eta_\mathrm{A} \right)\right), \nonumber \\
\beta_{13} &=& -i k_x k_y \omega c_\mathrm{A}^2 c_\mathrm{s}^3
   \left(c_\mathrm{A}^2 \chi_\mathrm{G}+ \eta_\mathrm{A}\right), \nonumber \\
\beta_{14} &=& 0, \nonumber \\
\beta_{21} &=& c_\mathrm{A}^2 c_\mathrm{s} \left[i \omega \left(-c_\mathrm{s}^2
   \left(k_x^2+k_y^2\right) \left(k_x^2 \eta_\mathrm{A} -\left(k_x^2+k_y^2
   \right) \eta-i \omega \right) \right.  \right.\nonumber \\
  &+& \left. \left.k_x^2 \left(\omega^2 \eta_\mathrm{A} +i g k_y
   \eta_\mathrm{H} \right) -\omega^2 \left(k_x^2+k_y^2\right) \eta \right)
   \right. \nonumber \\
  &+& \left. k_x^2 c_\mathrm{A}^2\left(-\omega^2+c_\mathrm{s}^2 \left(k_x^2-
   i k_y^2 \omega \chi_\mathrm{G} \right)\right)+\omega^4\right], \nonumber \\
\beta_{22} &=& -i \omega  c_\mathrm{A}^2 c_\mathrm{s} \left[ g \chi_\mathrm{g1}
   \left(c_\mathrm{s}^2 \left(k_x^2+k_y^2\right)-\omega^2\right) +i k_x^2 k_y
   c_\mathrm{s}^2 \eta_\mathrm{H} \right], \nonumber \\
\beta_{23} &=& -i \omega  c_\mathrm{A}^2 c_\mathrm{s} \eta
   \left(c_\mathrm{s}^2 \left(k_x^2+k_y^2\right)-\omega^2\right), \nonumber \\
\beta_{24} &=& 0, \nonumber \\
\beta_{31} &=& k_x \omega  c_\mathrm{A}^2 c_\mathrm{s} \left[ g k_y
   \chi_\mathrm{g1} \left(-k_x^2 c_\mathrm{A}^2+c_\mathrm{s}^2 \left(k_x^2+
   k_y^2\right)-\omega^2\right) \right. \nonumber \\
  &-& \left. g k_x^2 k_y \eta_\mathrm{A} +i
   \left(k_x^2 \eta_\mathrm{H} \left(c_\mathrm{s}^2 \left(k_x^2+k_y^2\right)
   -\omega^2\right)+g k_y \omega \right) \right. \nonumber \\
  &+& \left. g k_y \left(k_x^2+k_y^2\right)
   \eta \right], \nonumber \\
\beta_{32} &=& k_x k_y \omega  c_\mathrm{A}^2 c_\mathrm{s}^3 \left(k_x^2
   \left(c_\mathrm{A}^2 \chi_\mathrm{G}+ \eta_\mathrm{A} \right) -
   \left(k_x^2+k_y^2\right) \eta -i \omega \right), \nonumber \\
\beta_{33} &=&-g k_x k_y \omega c_\mathrm{A}^2 c_\mathrm{s} \eta, \nonumber \\
\beta_{34} &=& k_x k_y \omega c_\mathrm{A}^2 c_\mathrm{s}^3 \eta, \nonumber
\end{eqnarray}
\vspace{-1mm}
\noindent
\begin{eqnarray}
\gamma_{11} &=& i k_y \omega^2 c_\mathrm{s}^2 \eta_\mathrm{H}
   \left(c_\mathrm{s}^2 \left(k_x^2+k_y^2\right)-\omega^2\right), \nonumber \\
\gamma_{12} &=& \omega c_\mathrm{s}^2 \left[i c_\mathrm{A}^2 \left(\omega^2-
   c_\mathrm{s}^2 \left(k_x^2-i k_y^2 \omega  \chi_\mathrm{G} \right)\right)
   \right. \nonumber \\
  &+& \left. \omega \eta_\mathrm{A} \left(\omega^2-c_\mathrm{s}^2 \left(k_x^2+k_y^2
   \right)\right)+i g k_y \omega  \eta_\mathrm{H} \right], \nonumber \\
\gamma_{13} &=& \omega^2 c_\mathrm{s}^2 \left(-g \left(c_\mathrm{A}^2
   \chi_\mathrm{g1}+ \eta_\mathrm{A} \right)-i k_y c_\mathrm{s}^2
   \eta_\mathrm{H} \right),\nonumber \\
\gamma_{14} &=& \omega ^2 c_\mathrm{s}^4 \left(c_\mathrm{A}^2 \chi_\mathrm{G}
   +\eta_\mathrm{A}\right), \nonumber \\
\gamma_{15} &=& 0, \nonumber \\
\gamma_{21} &=& -i k_x \omega^2 c_\mathrm{s}^2 \eta_\mathrm{H}
   \left(c_\mathrm{s}^2 \left(k_x^2+k_y^2\right)-\omega^2\right), \nonumber \\
\gamma_{22} &=& k_x \omega  c_\mathrm{s}^2 \left(k_y c_\mathrm{A}^2
   c_\mathrm{s}^2 \left(\omega  \chi_\mathrm{G} -i\right)-i g \omega
   \eta_\mathrm{H} \right), \nonumber \\
\gamma_{23} &=& i k_x \omega^2 c_\mathrm{s}^4 \eta_\mathrm{H}, \nonumber \\
\gamma_{24} &=& \gamma_{25} = 0, \nonumber \\
\gamma_{31} &=& \omega c_\mathrm{s}^2 \left[ \omega^2-c_\mathrm{s}^2
   \left(k_x^2+k_y^2\right)\right) \left(-k_x^2 c_\mathrm{A}^2+\omega
   \left(i k_x^2 \eta_\mathrm{A} \right. \right. \nonumber \\
  &-& \left. \left. i \left(k_x^2+k_y^2\right) \eta +\omega
   \right)\right], \nonumber \\
\gamma_{32} &=& -i g \omega^2 c_\mathrm{s}^2 \left(k_x^2 \left(c_\mathrm{A}^2
   \chi_\mathrm{g1}+ \eta_\mathrm{A} \right)- \left(k_x^2+k_y^2\right) \eta
   -i \omega \right), \nonumber \\
\gamma_{33} &=& i \omega^2 c_\mathrm{s}^2 \left[ \omega^2 \eta +c_\mathrm{s}^2
   \left(k_x^2 \left(c_\mathrm{A}^2 \chi_\mathrm{G} +\eta_\mathrm{A} \right)
   -2 \left(k_x^2+k_y^2\right) \eta \right. \right. \nonumber \\
  &-& \left. \left. i\omega \right)\right], \nonumber \\
\gamma_{34} &=& -i g \omega^2 c_\mathrm{s}^2 \eta, \nonumber \\
\gamma_{35} &=& i \omega^2 c_\mathrm{s}^4 \eta.
\end{eqnarray}

The solution of the system of differential equations with constant coefficients
can be obtaining by a combination of exponentials in the form
\begin{equation}
b_x=A_1 \, e^{\lambda z}, \,\, b_y=A_2 \, e^{\lambda z}, \,\,
v_z=A_3 \, e^{\lambda z},
\label{acoefs}
\end{equation}

\noindent
with $A_1$, $A_2$ and $A_3$ arbitrary coefficients. This type of solution
leads to the system of algebraic equations
\begin{equation} \label{indicial_sist}
A_1 \sum_{j=1}^4 \alpha_{ij} \lambda^{j-1} + A_2 \sum_{j=1}^4 \beta_{ij}
 \lambda^{j-1} + A_3 \sum_{j=1}^5 \gamma_{ij} \lambda^{j-1} =0,
\end{equation}

\noindent
whose non-vanishing solutions are only obtained if the determinant of the
system is zero. This gives the indicial equation, which turns out to be a
$8^\mathrm{th}$ order algebraic equation in $\lambda$. There is no
simple way of expressing the solutions of this equation in terms of the
coefficients of the system, but for the range of parameters under
considerations, there are always four solutions with $\mathrm{Re}[\lambda]
\ge0$ and four with $\mathrm{Re}[\lambda] \le 0$, and can be further grouped in
pairs with $\mathrm{Re}[\lambda_1]\approx -\mathrm{Re}[\lambda_2]$ and
$\mathrm{Im}[\lambda_1]\approx \mathrm{Im}[\lambda_2]$, provided the diffusion
coefficients are small. It is easy to check that if all the diffusion
coefficients except the ambipolar one are set to zero the 8-th order equation
becomes Eq.~(\ref{indicial}), so two of the solutions are $\lambda \approx
m_1^{(1)}$ and $\lambda \approx m_2^{(1)}$ (and similar expression in the upper
zone), provided the diffusion terms are small, while the other two have much
larger real part than these two, representing exponentials that decay very fast
from the boundary. Finally, for each solution of the indicial equation we find a
relation between the $A$-coefficients in Eq.~(\ref{acoefs}), so for each value
of $\lambda$ all the perturbed velocity and magnetic field components can be
related to $b_x$, for example.

\end{appendix}

\bibliographystyle{aa}
\bibliography{ms.bbl}

\end{document}